\documentstyle[aps,manuscript]{revtex}
\newcommand{\equ}[1]{~Eq.~(\ref{#1})}
\newcommand{\Tr}{{\rm Tr\,}}
\newcommand{\vev}[1]{\left\langle #1 \right\rangle}
\begin{document}
\preprint{NYU-TH/98/05/21}
\draft

\title{Equivariant Gauge Fixing of SU(2) Lattice Gauge Theory}
\author{Martin Schaden}
\address{New York University, Physics Department\protect\\
 4 Washington Place, New York, New York 10003}
\date{\today}
\maketitle

\begin{abstract}
I construct  a  Lattice Gauge Theory (LGT) with
discrete $Z_2$ structure group and an equivariant BRST symmetry that
 is physically equivalent to the standard
$SU(2)$-LGT.  The measure of this $Z_2$-LGT is invariant under {\it
all} the discrete  symmetries of the lattice and its partition
function does not vanish. The Topological Lattice Theories
(TLT) that localize on the moduli spaces  are explicitly 
constructed and their BRST symmetry is exhibited.
The ghosts of the $Z_2$-invariant local  LGT are 
integrated in favor of a nonlocal bosonic measure. 
In addition to the $SU(2)$ link variables and
the coupling $g^2$, this effective bosonic measure
also depends on an auxiliary gauge invariant site variable of 
canonical dimension two and on a gauge parameter $\alpha$.  
The relation between
the expectation value of the auxiliary field, the gauge parameter
$\alpha$ and the lattice spacing $a$ is obtained to lowest order in
the loop expansion. In four dimensions and the critical limit this
expectation value is a
physical scale proportional to $\Lambda_L$ in the gauge $\alpha=g^2
(11-n_f)/24+ O(g^4)$. Implications for the  loop expansion of
observables in such a critical gauge are discussed.   
\end{abstract}


\section{Introduction}
Euclidean Lattice Gauge Theory (LGT) is the only known rigorous
non-perturbative definition of a non-abelian gauge 
theory. In the vicinity of a second order phase transition for a critical value
of the couplings, the LGT can be interpreted as
a regularization of a continuum quantum field theory in Euclidean
space-time.  Apart from  numerical simulations, such models also provide
a mathematically rigorous foundation for various non-perturbative
field theoretic ideas. These statistical models however also have
peculiarities of their own that  
have no analog in other regularizations of a quantum field theory. 

The discrete lattice by construction is devoid of any notion of
``smoothness'' and it is difficult to study 
effects related to topological
characteristics of the continuum gauge group. The ``gauge-group'' of
a LGT is simply
\begin{equation}
{\cal G}=\otimes_{sites} G_i \ ,
\label{latgroup}
\end{equation}
where the group $G_i$ at the $i$-th site is
isomorphic to the compact structure group $G$. 
Only the vanishing fraction of  lattice gauge transformations that
satisfy a Sobolev norm apparently correspond to continuum gauge
transformations\cite{De91} in the critical limit. For
lattice perturbation theory  and a continuum 
interpretation of the lattice model it is thus desirable to reduce the
rather large symmetry of the LGT to a more manageable level. This however
has to be done without altering physical observables of the model. The
procedure is (as we will see somewhat misleadingly) known as gauge fixing. 
One hopes that gauge fixing the lattice model would
help disentangle lattice gauge artifacts from the physically relevant
continuum dynamics.  The wild ``gauge''
group of the lattice preferably should be tamed in a fashion that assures
a smooth thermodynamic and critical limit of the physically equivalent
gauge fixed lattice model. In analogy with covariant gauges for the
continuum theory that preserve all the isometries of a space-time
manifold, a  gauge fixing procedure that preserves {\it all} the
(discrete) symmetries of a periodic lattice  will also be called 
``covariant'' in the following. While it is relatively simple to reduce the
gauge group of a LGT (by say ``fixing'' a maximal tree), it is
apparently not entirely trivial to obtain a {\it covariantly} gauge fixed
lattice measure that is normalizable\cite{Sh84,Ne87}.

In continuum perturbation theory, the method of choice for covariant
gauge fixing is BRST-quantization. Such gauges necessarily\cite{Si78} have a
Gribov-ambiguity \cite{Gr78}, i.e. an orbit generally crosses  the (covariant) gauge 
fixing surface more than once (and some orbits approach this
surface tangentially). Although apparently of little
relevance for an 
asymptotic perturbative expansion this ambiguity does concern the
non-perturbative validity of the gauge-fixed model. 
In the context of
Chern-Simons theory it was even recently shown that a correct
treatment of the generic gauge zero modes of degenerate 
background connections is essential for obtaining the (non-trivial)
asymptotic expansion of the model\cite{Ad98}. 

A valid non-perturbative
definition of the gauge fixed model is also of importance for the lattice.
It has been pointed out\cite{Hi79} that conventional BRST-invariant
Landau-gauge in fact counts the intersections of the orbit with a sign
that depends on the direction in which the oriented gauge fixing
surface is crossed -- the ``Gribov-ambiguity'' in this case would not
pose an obstruction to covariant gauge fixing as long as the degree of
this map does not 
vanish. Quite generally the degree of this
map however is zero for a covariantly gauge fixed LGT\cite{Sh84}. 

For continuum gauge theories the gauge fixing procedure was
recently seen to be equivalent to the construction of a 
Topological Quantum Field Theory (TQFT) on the 
gauge group\cite{Ba98}. It turns out that the partition
function of this TQFT is usually proportional to the generalized Euler
characteristic of the gauge group manifold and thus proportional
to the ``degree of the map'' of Sharpe. 
 The TQFT construction shows that
it is a topological characteristic of the {\it gauge group} that
determines whether or not the gauge-fixed theory makes sense
non-perturbatively. It allows one to continuously deform the orbit
and thus enables one to handle orbits that are {\it on} a Gribov
horizon. One can also  show that the
partition function of conventional covariantly gauge fixed continuum
models on compact space-time indeed vanishes non-perturbatively. 
The very construction of a
TQFT however allows one to address and solve these
problems\cite{Ba98,Ba96}.  
We will see that the method is also a powerful tool in the
construction of a physically equivalent and covariantly gauge-fixed LGT.  

The quest for a lattice analog of the elegant BRST-formalism of
continuum gauge theories has been elusive. Neuberger\cite{Ne86}
first formulated the analog of the conventional continuum
BRST-algebra for the lattice but subsequently  proved that the
partition function of a gauge-fixed lattice theory with this
BRST symmetry is not normalizable\cite{Ne87}. His proof is based on
particular properties of the BRST-algebra that do not hold for
the equivariant BRST construction we will consider below. For the
special case of certain covariant gauge fixings on the lattice,
Sharpe\cite{Sh84} had shown that the degree of the map is zero -- and
that the partition function of the gauge-fixed lattice theory
therefore vanishes due to the mutual cancellation of contributions
from different Gribov copies. His proof however appeared to depend on
the details of the gauge fixing and raises the question whether some other
covariantly gauge-fixed lattice action can be found. Sharpe proposed
several models whose partition functions do not vanish. In the
naive continuum limit some of them correspond to covariantly gauge
fixed actions.  These local lattice actions however break
some of the symmetries of a periodic lattice. Determining the
corresponding continuum model in this case requires a somewhat naive
extrapolation.  

I will translate the recent developments in continuum
BRST-quantization to the mathematically more rigorous setting of
LGT's on finite lattices. I use an equivariant BRST construction 
to reduce the gauge group of an $SU(2)$-LGT to a
physically equivalent abelian $U(1)$-LGT in section~II. [The
generalization of the 
procedure to other lattice gauge groups ${\cal G}$ and subgroups
${\cal H}\subset {\cal G}$ is relatively straightforward. The essential
point is to use a subgroup ${\cal H}$ for which the Euler characteristic
of the coset manifold $\chi({\cal G}/{\cal H})\neq 0$.]
In section~III I  examine the corresponding topological lattice theory
(TLT) and show  that it is a 
constant on the orbit space. The value of this constant
is explicitly computed in section~IV at the trivial link configuration
$U=1$.  I verify that the partition function of the TLT is  indeed
proportional to 
the Euler character $\chi({\cal G}/{\cal H}=(SU(2)/U(1))^N)=2^N\neq 0$
and therefore normalizable.  

This first step reduces the problem of
constructing a covariant and BRST-invariant gauge-fixed LGT to that of 
BRST-invariant gauge fixing of an U(1)-LGT. 
In section~V the presence
of local fields that are {\it charged} under the abelian group is utilized 
to build  a TLT that also fixes the residual abelian
invariance. The partition function of this TLT is shown to be
proportional to the number of connected components of the $U(1)$
gauge-group and is thus normalizable.   
One thus obtains  a local and
``lattice-covariant'' $Z_2$-LGT that is physically equivalent to the
original $SU(2)$-LGT. The loop expansion of this $Z_2$-LGT is examined
in section~VI.  I show that  
the measure is maximal at certain discrete pure gauge configurations
{\it and} a non-vanishing constant configuration
$\rho_i=\bar\rho(\alpha)$ of an auxiliary (gauge invariant) bosonic
field. The unique maximum of this bosonic measure
is determined  in the thermodynamic limit of a four dimensional  lattice.

\section{Equivariant BRST: Gauge-Fixing a SU(2)-LGT to a U(1)-LGT} 
Consider a $D$-dimensional LGT with an $SU(2)$ gauge
group and for simplicity  assume that the $SU(2)$-LGT is described by 
a local action  $S_{\rm inv.}[U]$ which depends only on the link
variables $U_{ji}^{\dagger}= U_{ij}\in SU(2)$. The generalization to the case
with matter fields is straightforward.  The invariance of the measure
with respect to the lattice gauge group\equ{latgroup} implies that
\begin{equation}
S_{\rm inv.}[U]=S_{\rm inv.}[U^{g\in{\cal G}}], \quad {\rm with}\quad U_{ij}^g=
g_i U_{ij} g_j^\dagger\ ,\ \ g_i\in SU(2)\ .
\label{transform}
\end{equation}

In this section we reduce the gauge invariance
of the LGT to the abelian subgroup 
\begin{equation}
{\cal H}=\otimes_{sites} U(1)\ ,
\label{latH}
\end{equation}
while preserving the locality of the measure and its invariance with
respect to the isometries of the lattice. The resulting model will
exhibit an equivariant BRST-symmetry and we will prove in sections~III
and~IV that it is  equivalent to the original $SU(2)$-LGT with
regard to physical observables.
 
The construction of the equivariant BRST
symmetry is analogous to the one in the continuum case\cite{Ba98}. Note
that an infinitesimal gauge transformation with
$g_i=1+\epsilon\theta_i+O(\epsilon^2)$ to order $\epsilon$ changes the
links by:
\begin{equation}
\Delta U_{ij}=\theta_i U_{ij}-U_{ij}\theta_j + O(\epsilon) \ .
\label{variation}
\end{equation}   
We accordingly define\cite{Ne86} the BRST-variation of $U_{ij}$ as
\begin{equation}
s U_{ij}= (c_i +\omega_i) U_{ij}-U_{ij}(c_j +\omega_j)\ ,\ \
\omega_i\in u(1)
\label{sU}
\end{equation}
where $c_i$ and $\omega_i$ are Lie-algebra valued Grassmannian site
variables. The reason for the apparently redundant  introduction of
{\it two} ghosts $c_i$ and $\omega_i$ instead of one for their sum is
that we can thus {\it specify} the action of one of
these ghosts and eventually decompose the Lie-algebra. For the case at
hand, we  take $\omega$ to be the ghost associated with the
generator of the $U(1)$ subgroup. Since our gauge fixing condition
will be $U(1)$-invariant, it is possible to arrange matters so
that the BRST-invariant  action of the physically equivalent $U(1)$-LGT does
not depend on the $\omega$ ghost. Requiring that the
BRST-variation be nilpotent, $s^2=0$, \equ{sU} implies
\begin{equation}
s c_i + s \omega_i= (c_i + \omega_i)^2 = c_i^2 + [\omega_i, c_i] +
\omega_i^2\ .
\label{scw}
\end{equation}  
Here $[\cdot,\cdot]$ is the commutator graded by the ghost
number. One satisfies\equ{scw} by 
\begin{eqnarray}
s c_i&=& c_i^2 + [\omega_i, c_i] +\phi_i\ ,\ \ \phi_i\in u(1)\nonumber\\
s \omega_i&=& \omega_i^2 -\phi_i =-\phi_i\nonumber\\
s \phi_i &=& [\omega,\phi_i] =0\ ,
\label{sc}
\end{eqnarray}
where the ghost number~$2$ field $\phi$ is introduced for the
following  reasons of
consistency.  Since the $\omega_i$ are in the Cartan sub-algebra $u(1)$ of
$su(2)$, we can without loss of generality demand that the $c_i$ 
span  the remaining two generators of the Lie-algebra. The necessary
Lagrange multiplier fields that implement this constraint will be
introduced below. Consistency however then requires that the component
in the Cartan sub-algebra of $s c_i$ also vanish. Since $c_i^2$
generally will (only) have a component in the Cartan sub-algebra, we can
satisfy this requirement only by introducing an additional field
$\phi\in u(1)$. Note that it is sufficient that $\phi$ take values in
the Cartan sub-algebra and that $\omega_i$ generates $U(1)$ transformations of
$c_i$ and $\phi_i$. Since the subgroup generated by $\omega$ in our
case is abelian, the BRST-variation of $\omega$ and $\phi$ simplify
in\equ{sc}.  In general, the equivariant BRST-construction
above can be employed to reduce any group ${\cal G}$ to a subgroup
${\cal H}\subset{\cal G}$ also for non-abelian ${\cal H}$. In\cite{Ba98}
a similar construction was for instance used to factor the global
gauge transformations of the continuum gauge theory. 

To complete the equivariant BRST construction one introduces 
Lagrange multiplier fields as BRST-doublets  that enforce the
constraints. For the gauge condition 
we require a Nakanishi-Lautrup field $b_i$ of vanishing ghost number. It
is part of the doublet
\begin{equation}
s \bar c_i=[\omega_i,\bar c_i] + b_i\ ,\ s
b_i=[\omega_i,b_i]-[\phi_i,\bar c_i]
\label{scbar}
\end{equation}
Note that the anti-ghost $\bar c_i$ here transforms under the $U(1)$. 
This is a natural consequence
of\equ{sU} -- we cannot take $\bar c_i$ to be neutral under $U(1)$,
because the BRST-invariant action we intend to construct would
otherwise be $\omega$-dependent. The BRST transformation of the
$b$-field is then given by the nil-potency of $s$. Note that the non-trivial
transformation of the Nakanishi-Lautrup 
field $b$ in\equ{scbar} in the present context invalidates Neubergers
proof\cite{Ne87} that the partition  function of a BRST-invariant  
lattice model is not normalizable.  To impose that
the components in the Cartan sub-algebra of 
$c, s c$ and $\bar c, s \bar c$ vanish, we need two more
doublets. The fields of these doublets  take values in the Cartan
sub-algebra only and therefore have the simple transformations
\begin{eqnarray}
s \bar\sigma=\sigma\ &,&\ s\sigma=0\ ,\ \ \bar\sigma,\sigma\in u(1)\nonumber\\
s \bar\gamma=\gamma\ &,&\ s\gamma=0\ ,\ \ \bar\gamma,\gamma\in u(1)
\label{ssg}
\end{eqnarray}

The construction of the partially gauge fixed action is completed by
specifying a local gauge fixing 
function $F_i[U]$ on the lattice configuration. A
sensible gauge fixing of the $SU(2)$ LGT to a $U(1)$ structure group has
to satisfy some non-trivial conditions. For any link configuration $U$
of the lattice there should at least be
one solution $g\in {\cal G}$ of
\begin{equation}
F_i[U^g]=0\ .
\label{cond1}
\end{equation} 
A $U(1)$-invariant gauge fixing furthermore requires that\equ{cond1}
be $U(1)$ invariant, that is 
\begin{equation}
F_i[U]=0 \Rightarrow F_i[U^h]=0, \forall h\in {\cal H}\subset {\cal G}
\label{cond2}  
\end{equation}
It is easy to see that\equ{cond1} always has a solution if the gauge
fixing function $F_i[U]$ is the Lie-derivative of a bounded Morse
potential $V[U]$
\begin{equation}
\sum_{i}\Tr\theta_i F_i[U]= \Delta V[U]\ ,
\label{varmorse}
\end{equation}
because \equ{cond1} then is the statement that 
$V[U^g]$, considered as a function of $g\in{\cal G}$ for fixed link
configuration $U$, has at least one extremum. This is certainly the
case for bounded $V[U]$.  \equ{cond2} is furthermore automatically
satisfied if the Morse potential is $U(1)$ invariant, i.e. 
\begin{equation}
V[U^h]=V[U],\ \forall h\in{\cal H}\ .
\label{invarV}
\end{equation}
To have a ``lattice-covariant'' gauge fixing we pick a local Morse
potential $V[U]$ that 
is a scalar under the action of the lattice group. The
simplest non-trivial Morse potential satisfying all these requirements
for the problem at hand is
\begin{equation}
V[U]=\sum_{\rm links}|\Tr \tau_+ U_{ij}|^2\ .
\label{defV}
\end{equation}
Here  $\tau_+=\tau_-^\dagger$ and $\tau_0$ are the $su(2)$ matrices of the
fundamental representation 
\begin{equation}
\tau_+=\left(\begin{array}{cc} 0 & 1\\ 0 & 0\end{array}\right)\ ,\ \
\tau_-=\left(\begin{array}{cc} 0 & 0\\ 1 & 0\end{array}\right)\ ,\ \
\tau_0=\frac{1}{2}\left(\begin{array}{cc} 1 & 0\\ 0 &
-1\end{array}\right)\ ,
\label{taus}
\end{equation}
with the commutation relations
\begin{equation}
[\tau_+,\tau_-]= 2 \tau_0\ ,\ \ [\tau_0,\tau_\pm]=\pm \tau_\pm\ .
\label{commutators}
\end{equation}
The potential\equ{defV} is  bounded below and on any {\it
finite} lattice is also bounded above. From\equ{variation} and the
definition\equ{varmorse} of the corresponding gauge fixing
function $F_i[U]$ one obtains 
\begin{eqnarray}
F_i[U]&=&\sum_{j\sim i} U_{ij}\tau_+(\Tr U_{ij}^\dagger\tau_-) - \tau_+
U_{ij}^\dagger (\Tr U_{ij}\tau_-)\nonumber\\
&&\qquad + U_{ij}\tau_-(\Tr
U_{ij}^\dagger\tau_+) - \tau_- U_{ij}^\dagger (\Tr U_{ij}\tau_+)\ .
\label{F}
\end{eqnarray}
Note that the gauge fixing function\equ{F} is anti-hermitian and  for a
particular site $i$ involves only  the links to the $2D$ adjacent
sites.  With the
$su(2)$~Lie-algebra\equ{commutators} one verifies that 
\begin{equation}
\Tr \tau_0 F_i[U] =0\ ,\ \ \forall U,
\label{direction}
\end{equation}
on any site $i$. This is a consequence of the
$U(1)$-invariance of the Morse potential\equ{defV}. To
construct the action we also need the BRST-variation of
$F_i[U]$. Because $\omega_i$ only has a component in
$\tau_0$-direction it is of the form 
\begin{equation}
s F_i[U]=[\omega_i, F_i[U]\,]+ M_i[U,c]\ ,
\label{sF}
\end{equation}
with
\begin{eqnarray}
M_i[U,c]&=&\sum_{j\sim i} (c_i U_{ij}\tau_+-U_{ij} c_j\tau_+)(\Tr
U_{ij}^\dagger\tau_-) \nonumber\\
& & +(\tau_+ U_{ij}^\dagger c_i -\tau_+ c_j
U_{ij}^\dagger)(\Tr U_{ij}\tau_-)\nonumber\\
& & +(U_{ij}\tau_+)(\Tr U_{ij}^\dagger
(\tau_- c_j -c_i\tau_-))\nonumber\\
& &+(\tau_+U_{ij}^\dagger)(\Tr U_{ij}
(c_j \tau_- -\tau_- c_i)) + (\tau_+\leftrightarrow \tau_-)\ . 
\label{defM}
\end{eqnarray}
Using a particular parameterization for the $SU(2)$ link variables, 
the  intimidating expressions \equ{F} and \equ{defM} are  
simplified in Appendix~A. For most of the following it suffices to
note that $M_i$ only involves links attached to the site $i$ and 
is linear in the ghost field $c$.

The action of the partially gauge fixed LGT is a local functional in the
equivariant cohomology of the BRST-symmetry we have defined. It is
thus of the form 
\begin{equation}
S=S_{\rm inv.} + S_{GF}\ ,\ \ {\rm with}\ S_{GF}=s W_{GF}\ ,
\label{action}
\end{equation}
where $W_{GF}$ is a local lattice action of ghost number~1 that is
$U(1)$ invariant and does not involve the $\omega$-ghost. The
restriction to operators that are relevant in the critical limit 
imposes additional  constraints on $W_{GF}$. The
most general relevant  $W_{GF}$ for  the $SU(2)$ model is,
\begin{equation}
W_{GF}=\sum_i \Tr\left[ \bar c_i F_i[U] +\frac{\alpha}{2}
\bar c_i b_i+ \beta {\bar c}_i^2 c_i\right] +\bar\gamma_i\Tr\tau_0\bar c_i +\bar\sigma_i\Tr\tau_0 c_i\ .
\label{WGF}
\end{equation}
This gauge fixing functional depends on two gauge parameters $\alpha$ and
$\beta$. Using equations\equ{sc},\equ{scbar},\equ{ssg} and\equ{sF} one
finds 
\begin{eqnarray}
S_{GF}&=& s W_{GF}=\sum_i\Tr\left[ b_i F_i[U] -\bar c_i M_i[U,c]
+\frac{\alpha}{2} b_i^2+ \beta b_i[\bar c_i,c_i] +\beta {\bar c}_i^2
c_i^2 \right]\nonumber\\
& & + (\beta-\alpha) \phi^0_i\Tr\tau_0 {\bar c}_i^2
+\frac{1}{2}\bar\sigma_i\phi^0_i +\bar\sigma_i\Tr\tau_0
c_i^2\nonumber\\
& &+\gamma_i\Tr\tau_0\bar c_i +\bar\gamma_i\Tr\tau_0 b_i
+\sigma_i\Tr\tau_0 c_i\ ,
\label{SGF}
\end{eqnarray} 
where use has been made of the fact that $\omega_i=\omega_i^0\tau_0$
and $\phi_i=\phi^0_i\tau_0$ are fields with values in the Cartan
sub-algebra only. If we only consider expectation values of
functionals that do not depend on $\phi,\bar\sigma,\gamma,\bar\gamma$
nor $\sigma$, these fields can be eliminated by their equations of motion.

The last three terms in\equ{SGF} enforce that $\bar c, c$ as well as
$b$ are orthogonal to the $\tau_0$-direction, i.e. they eliminate the
$U(1)$-neutral components of these fields. It follows that $\Tr
b_i[\bar c_i,c_i]=0$ and that $c_i^2$ as well as ${\bar c}_i^2$ only
have components in the $\tau_0$-direction. The equations of motion for
$\bar\sigma_i$ and $\phi_i^0$ then give rise to a quartic ghost
interaction which after a bit of algebra can be brought in the form
$(\alpha-\beta)\Tr {\bar c}_i^2 c_i^2$. These manipulations lead to 
a substantially simplified (on-shell) action   
\begin{equation}
S_{GF}^{\text{on shell}}=\sum_i\Tr\left[ b_i F_i[U] -\bar
c_i M_i[U,c] +\frac{\alpha}{2} b_i^2+ \alpha {\bar c}_i^2 c_i^2
\right]\ ,
\label{onshell}
\end{equation}
where $b_i,\bar c_i$ and $c_i$ only have components that are
charged under the $U(1)$. Note that the on-shell action\equ{onshell}
no longer depends on the gauge parameter $\beta$ of\equ{WGF}. The
equations of motion have removed this parameter in favor of
a quartic ghost interaction proportional to $\alpha$. There is a
quartic ghost interaction in any gauge $\alpha\neq
0$. It is a consequence of
the equivariant BRST construction and does not depend on the employed
gauge fixing function $F_i$.  As we will
see in the next section there is a good reason for this quartic 
ghost interaction. Let me  comment here that Landau gauge
with $\alpha=0$ is in a certain sense an {\it impossible} gauge on the
lattice that can only be perturbatively defined. A
non-perturbative definition of Landau gauge
would require finding {\it all} solutions to the constraint
$F_i[U^g]=0$ {\it exactly} for any configuration $U$. 
The problem is equivalent to  finding all extrema
of the Morse function $V[U]$ {\it exactly} -- clearly an impossible
task for any algorithm. Unlike the constraints on the fields that
we solved to arrive at\equ{onshell}, the condition
$F_i[U]=0$ is non-local and {\it cannot} be {\it solved} analytically for large
lattices. Perhaps more important, the  
error in the estimation of an extremum of $V[U]$ to any finite
numerical accuracy can be shown to grow  rapidly with the number of
lattice sites. In terms of the Lattice Renormalization Group,
$\alpha=0$ is  an {\it unstable} fixed point.    

For the proof of the next two sections that the 
partially gauge fixed lattice theory is physically equivalent to 
the $SU(2)$-LGT, it is useful to also eliminate the charged
Nakanishi-Lautrup field $b_i$.  Due to\equ{direction} one obtains the
effective gauge fixing action
\begin{equation}
S_{GF}^{\rm eff}=\sum_i \Tr\left[ -\frac{1}{2\alpha} F_i[U] F_i[U]
-\bar c_i M_i[U,c] +\alpha{\bar c}_i^2 c_i^2 \right]\ ,
\label{effSGF}
\end{equation}
where $c_i$ and $\bar c_i$ have only components that are charged under
the $U(1)$.
Numerical integration of  Grassmannian variables is not possible
and the local action\equ{effSGF}  so far is a  mathematical
construct. To explicitly perform the Grassmann integrals,\equ{effSGF}
would have to be bilinear in the ghosts $c$ and $\bar c$. Since $M_i[U,c]$
given by\equ{defM} is linear in the ghost $c$ this objective 
is achieved by introducing an auxiliary site-variable $\rho_i$
with vanishing ghost number to  {\it linearize}  the quartic ghost
interaction. The action
\begin{equation}
S_{GF}^{\rm linear}=\sum_i \Tr\left[ -\frac{1}{2\alpha}
F_i[U] F_i[U] 
-\bar c_i M_i[U,c] - \rho_i \tau_0 [{\bar c}_i, c_i]\
\right] +\frac{1}{4\alpha} \rho_i^2\ ,
\label{linSGF}
\end{equation}  
is equivalent to\equ{effSGF} upon using the equation of
motion of $\rho_i$ and the fact that $[\bar c_i, c_i], c_i^2, \bar
c_i^2$ all are in the Cartan sub-algebra. The Grassmann integrals of 
the partition function over $c_i$ and $\bar c_i$ can now be performed
analytically and give the determinant of a matrix  that depends on the link
configuration $U$ {\it and} the auxiliary field $\rho$. 
In a numerical simulation of the partially
gauge fixed LGT the Gaussian average of this determinant over $\rho$
determines the measure for the link variables.  

Before proving that the partition function of the
partially gauge fixed LGT does not vanish, note that the gauge fixed
action $S^{\rm eff}=S_{\rm inv.} + S_{GF}^{\rm eff}$ is invariant under the
following relatively simple on-shell BRST symmetry $\tilde s$:
\begin{equation}
\tilde s U_{ij}=c_i U_{ij} - U_{ij} c_j\ ,\ \ \tilde s c_i =0\ , \ \ 
\tilde s \bar c_i= -\frac{1}{\alpha} F_i[U]\ ,
\label{onshellbrs}
\end{equation}
where the ghost fields satisfy the constraints
\begin{equation}
\Tr \tau_0 c_i=\Tr \tau_0 \bar c_i=0\ ,
\label{charged}
\end{equation}
and $F_i[U]$ is given by\equ{F}. Note that the constraint\equ{charged}
on $\bar c_i$ is consistent with\equ{onshellbrs} due
to\equ{direction}. Furthermore $\tilde s$ is on-shell nilpotent 
on functions that are invariant with respect to the $U(1)$ gauge
group. Thus 
\begin{equation}
{\tilde s}^2 U_{ij}=c_i^2 U_{ij}-U_{ij} c_j^2
\label{s2U}
\end{equation}
effects an infinitesimal $U(1)$ gauge transformation  generated by
$c^2\propto \tau_0$. Using the equation of motion for $\bar c_i$
\begin{equation}
M_i[U,c]=\alpha [\bar c_i, c_i^2]\ ,
\label{EMcbar}
\end{equation}
we have that  
\begin{equation}
{\tilde s}^2 \bar c_i=-\frac{1}{\alpha} M_i[U,c]\simeq [c_i^2, \bar c_i]\ ,
\label{s2cbar}
\end{equation}
and thus on-shell is equivalent to an infinitesimal $U(1)$ gauge transformation
generated by $c^2$. We similarly obtain using\equ{F} and\equ{s2U} that
\begin{equation}
{\tilde s}^2 F_i[U]=\tilde s M_i[U,c]=[c_i^2, F_i[U]]\ .
\label{s2F}
\end{equation}
The BRST-symmetry  $\tilde s$ thus defines an
{\it equivariant} cohomology on the (graded) Grassmann algebra of the
set of $U(1)$-invariant functions 
\begin{equation}
{\cal B}:=\{A[U,c]: A[U^h, c^h]=A[U,c]\  \forall
h\in{\cal H}\} \ , 
\label{set}
\end{equation} 
 of the link variables and ghost field $c$. The
nontrivial observables of the partially gauge fixed LGT  is
the equivariant cohomology $\Sigma$,
\begin{equation}
\Sigma:=\{O\in {\cal B}: \tilde s O=0, O\neq \tilde s E,
\forall E\in {\cal B}\}\ .
\label{cohomology}
\end{equation}    
The functions in $\Sigma$ with vanishing ghost number are 
the gauge-invariant functions of the links only,
i.e. Wilson loops and their (linked) products. The 
physical observables of the original $SU(2)$-LGT  thus constitute the
sector with vanishing ghost number  of  
the equivariant cohomology $\Sigma$.

\section{The Topological Lattice Theory (TLT)}
I still have to show that the expectation value of a physical
observable $O[U]\in\Sigma$
\begin{equation}
\vev{O[U]}:=\int \prod_{links} dU_{ij}\prod_{sites} d^2c_id^2\bar c_i O[U]
\,\exp \{-S^{\rm eff}[U,c,\bar c,\alpha]\}\ , 
\label{expect}
\end{equation}
with the action
\begin{equation}
S^{\rm eff}[U,c,\bar c,\alpha]=S_{\rm inv.}[U] + S_{GF}^{\rm
eff}[U,c,\bar c,\alpha]\ , 
\label{effaction}
\end{equation}
up to an overall (non-vanishing) normalization ${\cal N}(\alpha)$ 
is the expectation value of the observable in the original LGT with the
gauge-invariant measure. We thus wish to show that
\begin{equation}
\vev{O[U]}={\cal N}(\alpha)\int \prod_{links} dU_{ij} O[U] \exp\left\{-S_{\rm
inv.}[U] \right\}=: \vev{O[U]}_{\rm inv.}\ ,
\label{want}
\end{equation}
for {\it all} physical observables $O[U]$ and any {\it finite} lattice.

Since the volume of the $SU(2)$ lattice gauge group of a finite
lattice is a finite non-vanishing constant, 
\begin{equation}
V_{\cal G}=\int \prod_{sites} dg_i <\infty\ ,
\label{groupvolume} 
\end{equation} 
we can multiply both sides of\equ{expect} by $V_{\cal G}$ and change
the  integration variables  
\begin{equation}
U_{ij}= U_{ij}^{\prime\,g}= g_i U_{ij}^\prime g_j^\dagger
\label{change}
\end{equation}
The Haar-measure $d U_{ij}= d U_{ij}^\prime $ as well as the gauge
invariant part of the lattice action and the observable $O[U]=O[U^\prime]$
are invariant under this (gauge) transformation and\equ{expect} becomes 
\begin{equation}
V_{\cal G}\vev{O[U]}=\int \prod_{links} dU_{ij} O[U] {\cal Z}[U,\alpha]
e^{-S_{\rm inv.}[U]}\ ,
\label{factorize}
\end{equation}
where 
\begin{equation} 
{\cal Z}[U,\alpha]=\int \prod_{sites} dg_i d^2c_i d^2\bar c_i e^{- S_{GF}^{\rm
eff}[U^g,c,\bar c;\alpha]}\ .
\label{TQFT}\
\end{equation}
Evidently  ${\cal Z}[U^g,\alpha]={\cal Z}[U,\alpha]$ 
is itself a gauge invariant observable. For
\equ{want} to hold for {\it all} observables $O[U]$, ${\cal
Z}[U,\alpha]$ must be a constant that does not depend on the link
configuration $U$ at all. We therefore have to show that\equ{TQFT} is a
non-vanishing constant on the configuration space, i.e. that the model
defined by the partition function\equ{TQFT} is a TLT.     

I will first determine the $\alpha$-dependence of ${\cal
Z}[U,\alpha]$ and then show that this partition function does not
depend on a continuous deformation of the configuration $U$.
The basis for these conclusions is that $S_{GF}^{\rm
eff}[U^g,c,\bar c;\alpha]$ is invariant with respect to a (on-shell
nilpotent) BRST-symmetry $\hat s$  defined on the variables as 
\begin{eqnarray}
\hat s U_{ij}=0 & &\nonumber\\
\hat s g_i=c_i g_i\ &,&\ \ \hat s g_i^\dagger=-g_i^\dagger c_i\nonumber\\
\hat s c_i=0 \ &,& \ \ \hat s \bar c_i=-\frac{1}{\alpha} F_i[U^g]\ .
\label{hats}
\end{eqnarray}
Note that the algebra\equ{hats} is very similar to the
BRST-algebra\equ{onshellbrs} but does not transform the link
configuration $U$. The third relation in\equ{hats} is a consequence of
the second and  $g_i g_i^\dagger=1$. 
The invariance of $S_{GF}^{\rm
eff}[U^g,c,\bar c;\alpha]$ follows immediately  
from the invariance of $S_{GF}^{\rm eff}[U,c,\bar c;\alpha]$ under $\tilde s$ 
and
\begin{equation}
\hat s (g_i U_{ij} g_j^\dagger)=c_ig_i U_{ij} g_j^\dagger-
g_i U_{ij} g_j^\dagger c_j=\left.\tilde s
U_{ij}\right|_{U_{ij}\rightarrow g_iU_{ij}g_j^\dagger}
\label{equival}
\end{equation}
Since $\hat s c=0$, the measure $dg_i d^2c_i
d^2\bar c_i$ in\equ{TQFT} is evidently $\hat s$-invariant if $dg_i$
is the Haar-measure of the structure group.

To simplify notation I define the (not normalized) expectation value
of any function $X$ of the fields $g,c,\bar c$ in the TLT  
\begin{equation}
\vev{X}_{U,\alpha}:=\int \prod_{sites} dg_i d^2c_i d^2\bar c_i X e^{-S_{GF}^{\rm
eff}[U^g,c,\bar c;\alpha]}\ . 
\label{expectU}
\end{equation}
The function $X$ can  itself depend parametrically on the
configuration $U$ and the gauge parameter 
$\alpha$. In this notation, ${\cal Z}$ of\equ{TQFT} is just
$\vev{1}_{U,\alpha}$.  Using\equ{effSGF}, the definition\equ{TQFT}
implies that
\begin{eqnarray} 
\alpha\frac{\partial}{\partial\alpha} {\cal
Z}[U,\alpha]&=&-\vev{\sum_{sites}\Tr\frac{1}{2\alpha} F_i[U^g]
F_i[U^g] +\alpha {\bar c}_i^2 c_i^2}_{U,\alpha}\nonumber\\
&=& \vev{\hat s \sum_{sites}\frac{1}{2}\Tr \bar c_i
F_i[U^g]}_{U,\alpha} +\vev{\sum_{sites}\Tr \frac{1}{2} \bar c_i
M_i[U^g,c]-\alpha{\bar
c}_i^2 c_i^2 }_{U,\alpha}\nonumber\\
&=& N Z[U,\alpha]\ ,
\label{alphavar}
\end{eqnarray}
for a lattice with $N$ sites. The last equality in\equ{alphavar} is a
consequence of the $\hat s$-invariance of $S_{GF}^{\rm eff}$ and the measure
and of the equation of motion\equ{EMcbar}. Due to\equ{alphavar}, 
\begin{equation}
\widetilde{\cal Z}[U]=\alpha^{-N} {\cal Z}[U,\alpha]\ ,
\label{constZ}
\end{equation}
is a gauge invariant functional of the configuration $U$ that does not
depend on $\alpha$.  

Similar reasoning shows that ${\cal Z}[U,\alpha]$ does not change
under a {\it continuous} 
deformation of the orbit. Since $\hat s U=0$, we  symbolically have
\begin{eqnarray}
\delta_U Z[U,\alpha]&=& \vev{\delta_U \sum_i \Tr\left
[ \frac{1}{2\alpha} F_i[U^g] F_i[U^g] +\bar c_i 
M_i[U^g,c]\right]}_{U,\alpha}\nonumber\\
&=&-\vev{\hat s\delta_U \sum_i \Tr
\frac{1}{2} \bar c_i F_i[U]}_{U,\alpha} +\vev{\delta_U\sum_i \Tr\left[\frac{1}{2} \bar c_i M_i[U^g,c]
-\alpha{\bar c}_i^2 
c_i^2\right]}_{U,\alpha}\nonumber\\
&=& 0\ ,
\label{const}
\end{eqnarray}
where the last equality again makes use of the equation of
motion\equ{EMcbar}. (In\equ{const} the variation $\delta_U$ of the
link variables of course respects $U_{ij}\in
SU(2)$.)  

The property\equ{const} that ${\cal Z}[U,\alpha]$ (and thus  also
$\widetilde{\cal Z}[U]$) is  constant on a
connected set of link configurations greatly simplifies our task. To 
determine the value of $\widetilde {\cal Z}[U]$ we need only consider a
particular link configuration in each connected sector of the orbit
space.  In a LGT {\it every} link configuration is
connected to the trivial one with $U_{ij}=1$ on all
links. Thus\equ{const} implies that $\widetilde{\cal Z}[U]$ is a
constant that does not depend on the link configuration. To show that
this constant does not {\it vanish}, it is sufficient that 
\begin{equation}
\widetilde{\cal Z}[U=1]\neq 0\ ,
\label{nonzero}
\end{equation}
for any finite lattice. \equ{const}
and\equ{constZ} together with\equ{factorize}  imply that the
expectation value\equ{expect} of {\it any} physical observable $O$  
in the partially gauge fixed LGT is proportional to the expectation
value of the same observable in the original $SU(2)$-LGT. The
proportionality constant furthermore does not depend on the observable
and does not vanish when \equ{nonzero} holds.

\equ{const} together with\equ{constZ} establish that the model
described by the partition function $\widetilde{\cal Z}[U]$ is the
lattice version of a TQFT (of Witten type) on the space ${\cal
G}/{\cal H}$. The partition function of this TLT is some
topological characteristic of the coset space. In the next section we will
explicitly demonstrate that $\tilde {\cal Z}$ is proportional to the
Euler characteristic $\chi({\cal G}/{\cal H})$. Since $\chi({\cal
G}/{\cal H})=\chi(\otimes_{sites} SU(2)/U(1)\simeq
S_2)=(\chi(S_2))^N=2^N\neq 0$, this will prove that
$\widetilde{\cal Z}$ indeed does not vanish. Note that the basic
reason for only partially gauge fixing the $SU(2)$-LGT using an
equivariant  BRST construction was that $\chi({\cal 
G})=0$  -- the partition function of a TLT that is proportional to
the Euler character of the compact lattice gauge group would have
vanished no matter what Morse potential one chooses.

\section{Semi-classical evaluation of $\widetilde{\cal Z}[U=1]$.}
Although multi-dimensional, a LGT is nevertheless only a
statistical mechanical system. Even more importantly, the variables of
this system are compact. Consequently the lattice action $S_{\rm
inv.}$, and also $V[U]$ 
defined by\equ{defV} are {\it bounded} functions for any finite
lattice. We are in the  fortunate position that almost
all requirements of Morse theory (which generally 
applies to compact spaces and bounded functions) are satisfied
for the TLT. At this point we could therefore simply cite the
Poincar\'e-Hopf Theorem and known results from topological
quantum mechanical models\cite{Bi91} to assert that the partition function
$\widetilde {\cal Z}[U]$ is proportional to the Euler characteristic
$\chi({\cal G}/{\cal H})$ of the manifold that is the domain of the
{\it bounded} Morse-function $V_U[g]$ 
\begin{equation}
V_U[g]= V[U^g]: {\cal G}/{\cal H}\rightarrow R\ ,
\label{mapping}
\end{equation}
when $V[U^g]$ is considered as a function of the gauge transformation
for fixed link configuration $U$. Since $\chi({\cal G}/{\cal
H})=2^N\neq 0$ this would prove our assertions. 

The TLT on the other hand is a sufficiently simple model for us 
to {\it explicitly} see these topological theorems at work. The
following computation of $\widetilde{\cal Z}[1]$ also shows which
``pure gauge'' configurations give a vanishing contribution to
$\widetilde{\cal Z}[1]$ in the limit $\alpha\rightarrow 0$ and which
don't. In section~VI this gives us greater certainty in the  evaluation
of correlation functions 
in the {\it critical limit} $g^2 \rightarrow 0$ of the gauge fixed
model since only a certain class of saddle points contributes in the
limit $\alpha\rightarrow 0$. In the course of the calculation we will
furthermore characterize {\it all } Gribov copies of the vacuum
configuration $U=1$ to the gauge condition $F_i[1^g]=0$. Perhaps the
most interesting aspect of the computation  is the important role of
the quartic ghost interaction in\equ{effSGF}. 

Using the result of the previous section that $\widetilde{\cal
Z}$ does not depend on the gauge parameter $\alpha$, we may choose
$\alpha$ sufficiently small for a saddle point approximation to the
integral\equ{TQFT} to be as accurate as we please. Although I will not
explicitly  compute the errors of the saddle point approximation, it
is quite obvious that the evaluation becomes exact in the limit
$\alpha\rightarrow 0$ for a lattice with $N<\infty$ sites, because 
$\sum_i \Tr F_i[1^g] F_i[1^g]$ in this case is a bounded function on a
{\it finite} dimensional space of gauge transformations.  

To compute
\begin{equation}
\widetilde {\cal Z}[1]=\lim_{\alpha\rightarrow 0_+} \widetilde{\cal
Z}[1]\ ,
\label{saddle}
\end{equation} 
with the action\equ{effSGF} in the definition\equ{TQFT} of ${\cal Z}$,
we need to consider {\it all} solutions $\tilde g$ to
the equations 
\begin{equation}
F_i[1^{\tilde g}]=0\ \ \forall\ {\rm
sites}\ i\ .
\label{extrema}
\end{equation}
Because $F_i[U]$ is the Lie-derivative\equ{varmorse} of the
Morse-potential\equ{defV}, \equ{extrema} in
principle requires us to determine all {\it extrema} of 
\begin{equation}
V[g]=V[U_{ij}=g_i g_j^\dagger]=\sum_{\rm links}|\Tr \tau_+ g_i
g_j^\dagger|^2\ ,
\label{Vg}
\end{equation}
in the space of lattice gauge transformations. By construction,
\equ{Vg} is invariant with respect to left-handed $U(1)$ gauge
transformations $h\in{\cal H}$,
\begin{equation}
V[hg]=V[g]\ \ \forall\ h\in{\cal H}\ .
\label{invVg}
\end{equation}
We can use the invariance\equ{invVg} to parameterize the $SU(2)/U(1)$
coset element $g_i$ at each site by only {\it two real} angles,
\begin{equation}
g_i=\left(\begin{array}{cc} \cos(\theta_i/2) & \sin(\theta_i/2)
e^{i\varphi_i} \\ -\sin(\theta_i/2) e^{-i\varphi_i} & \cos(\theta_i/2)
\end{array}\right)\ ,
\label{parameterg}
\end{equation}
with $\theta_i\in [0,\pi]$ and $\varphi_i\in [0,2\pi)$. We can always
choose $h_i\in U(1)$ to eliminate the phase in the 
diagonal elements of $g_i$. At $\theta_i=\pi$ the diagonal elements of
$g_i$ in\equ{parameterg} vanish and the phase of the off-diagonal
elements can be arbitrarily changed by an
$U(1)$-transformation. Identifying all the points  $(\theta=\pi,\varphi)$
we see that there is a  one-to-one correspondence 
\begin{equation}
g_i(\theta_i,\varphi_i)\in SU(2)/U(1) \leftrightarrow \hat
s_i\in S_2\ ,
\label{correspondence}
\end{equation}
between $g_i\in SU(2)/U(1)$ and unit ``spins'' $\hat s_i=(\sin\theta_i
\cos\phi_i, \sin\theta_i \sin\varphi_i, 
\cos\theta_i)$ describing a two-dimensional sphere. This is of course
just the statement that the coset manifold $SU(2)/U(1)\simeq S_2$.
Using the parameterization\equ{parameterg}, \equ{Vg} after a bit of 
algebra can be seen to be the energy of the Heisenberg model,
\begin{equation}
V[g]=\frac{1}{4}\sum_{i\sim j} (\hat s_i -\hat s_j)^2 \ .
\label{Vspin}
\end{equation}

The relation\equ{Vspin} helps to visualize and classify
the extrema of $V[g]$. $V[g]$ possesses  a {\it continuous global}
$SO(3)$ invariance corresponding to a 
coherent rotation of all the spins $\hat s_i$. The extrema of $V[g]$
are thus characterized by the subgroup of $SO(3)$ under which they
are invariant. There are only two kinds of extrema:
\begin{itemize}
\item[I)] extrema that are invariant under an $SO(2)$ subgroup of
$SO(3)$. In this case all the spins are collinear. There are {\it two}
zero modes associated with any extremum of this kind, corresponding to
the broken generators of the coset space $SO(3)/SO(2)$. These zero modes
correspond to infinitesimal global rotations of the collinear spins, whereas an
$SO(2)$-rotation along the axis of any particular spin does not 
change these extremal configurations. Thus type~I extrema fall in classes
$[\tilde g]_I$ modulo global rotations of all the
spins. One can select a unique representative of such a class by
specifying the direction of any particular spin. There is a one-to-one
correspondence between configurations in $[\tilde g]_I$ and 
points on a two-dimensional sphere $S_2$. Since all the spins are
collinear, there are exactly $2^{N-1}$ classes $[\tilde g]_I$ on a
lattice with $N$ sites. 
\item[II)] extrema that are {\it not} invariant under any continuous 
subgroup of $SO(3)$. In this case the spins are not {\it all}
collinear. An example of this kind of extrema are the solitons of the
1-dimensional periodic spin chain\cite{Je79}. By Goldstone's theorem
there are {\it three} zero modes corresponding to the generators of
$SO(3)$. Their action on a particular configuration can be
visualized as follows: two generators correspond to global rotations
of the extremal configuration. The third effects an infinitesimal
$SO(2)$-rotation of the configuration along the axis of a {\it
particular} spin. Thus 
type~II extrema fall into classes $[\tilde g]_{II}$ modulo
$SO(3)$ rotations. A particular representative of such
a class is selected by specifying the direction of one of the spins,
say $\hat s_0$  {\it and} the direction of $\hat s_0\times \hat s_j$,
where $s_j$ is a specific spin of the configuration that is {\it not}
collinear to $s_0$. There is thus a one-to-one correspondence between
configurations in $[\tilde g]_{II}$ and the points of a
3-dimensional sphere $S_3$. 
\end{itemize}
The above classification of the extrema of $V[g]$ is complete in the
sense that there are no other {\it continuous} symmetries relating 
extremal configurations.  

Expanding
$V[g]$ and $F_i[1^g]$ to quadratic-, respectively linear-, order near
an extremum $\tilde g$ one has  
\begin{eqnarray}
V[g=\tilde g e^\theta ]&=& V[\tilde g] -\sum_i \Tr
\theta_i^\dagger M_i[1^{\tilde g},\theta] + O(\theta^3)\nonumber\\
F_i[1^g]&=& M_i[1^{\tilde g},\theta] + O(\theta^2)\ .
\label{expand}
\end{eqnarray}
For the saddle point evaluation it is useful to expand in terms of
eigenvectors of the $2N$ linear equations 
\begin{equation}
M_i[1^{\tilde g},\phi^{(n)}]=\lambda_{(n)} \phi_i^{(n)}\ ,\ \
n=1,2,\dots,2N 
\label{eigeneq}
\end{equation}
where the eigenvalues
$\lambda^{(n)}$ and eigenvectors $\phi^{(n)}$ implicitly depend on the
extremum $\tilde g$. Since $V[g]$ in\equ{Vspin} is a globally
$SO(3)$ invariant real function of the spins,
\equ{expand} implies  that the eigenvalues $\lambda_{(n)}$ are real
and depend only on the class $[\tilde g]$ of the extremal
configuration (and not on the particular representative of that
class). Since the quadratic form in\equ{expand} is real, the
eigenvectors $\phi^{(n)}$ furthermore can be chosen to form a complete
orthonormal set with respect to the inner product
\begin{equation}
\vev{n|m}=\sum_{sites}\Tr
\phi_i^{(n)\dagger}\phi_i^{(m)}=\delta_{nm}\ .
\label{orthonormal}
\end{equation} 

In the vicinity of an extremal configuration $\tilde g$, 
the action $S_{GF}^{\rm eff}$ is of the form (using the expansions
\equ{expand}), 
\begin{equation}
S_{GF}^{\rm eff}[\alpha;{\tilde ge^\theta},c,\bar c]\sim
\sum_i \Tr\left[ \frac{1}{2\alpha} M_i^\dagger[1^{\tilde g},\theta]
M_i[1^{\tilde g},\theta] 
-\bar c_i M_i[1^{\tilde g},c] +\alpha{\bar c}_i^2 c_i^2
\right]\ ,
\label{extremeaction}
\end{equation}
up to terms of order $\theta^3$, respectively $\theta c\bar
c$. Since we omitted terms of order $\theta^3$ and $\theta\bar c c$ in the
expansion\equ{extremeaction}, retaining the quartic
ghost interaction could appear questionable. We will however soon see that 
the sole purpose of the quartic ghost interaction to leading order in
$\alpha$ is to absorb Grassmannian zero-modes. The neglected terms are
higher order variations of the Morse-potential and therefore do not
couple to the zero-modes. The {\it leading}
contribution in $\alpha$ can thus be calculated
using\equ{extremeaction}. Note also that $M_i[U,\theta]$ given
by\equ{defM} is anti-hermitian.  
 
To diagonalize the quadratic form  
in\equ{extremeaction} we expand $\theta, c$ and $\bar c$ in the
complete set of orthonormal eigenvectors of\equ{eigeneq} 
\begin{equation}
\theta_j=i\sum_n\xi^{(n)}\phi_j^{(n)}\,,\ c_j=\sum_n
c^{(n)}\phi_j^{(n)}\,,\ \bar c_j=\sum_n \bar
c^{(n)}\phi_j^{(n)\dagger}\ ,
\label{coeffs}
\end{equation}
with real coefficients $\xi^{(n)}$ and Grassmannian variables $c^{(n)},\bar
c^{(n)}$. In terms of these coefficients the action\equ{extremeaction}
in the vicinity of the extremum $\tilde g$ takes the form
\begin{equation}
S_{GF}^{\rm eff}[\tilde g,\alpha; \{\xi^{(n)},c^{(n)},\bar c^{(n)}\}]\sim
\sum_n\left[  \frac{1}{2\alpha} \xi^{(n)}\lambda^2_{(n)}\xi^{(n)}-\bar
c^{(n)} \lambda_{(n)} c^{(n)} +\alpha\sum_{klm} R_{klmn} {\bar
c}^{(k)} {\bar c}^{(l)} {c}^{(m)} {c}^{(n)}
\right]\ ,
\label{diagonal}
\end{equation} 
with
\begin{equation}
R_{klmn}=\sum_{sites}\Tr
\phi_i^{(k)\dagger}\phi_i^{(l)\dagger} \phi_i^{(m)}\phi_i^{(n)}\ .
\label{R}
\end{equation}
The change of basis\equ{coeffs} diagonalizes the quadratic part of the
action near an extremum. The remaining quartic ghost interaction
is irrelevant for the  semi-classical evaluation {\it except} for 
Grassmannian zero-modes that do not enter quadratically. 
 
As noted above, extrema of
type~I are characterized by  {\it two} zero-modes with vanishing
eigenvalues. I will denote these eigenvectors by
$\phi^{(1)},\phi^{(2)}$ in the following
($\lambda_{(1)}=\lambda_{(2)}=0$).  The $SO(2)$ symmetry of  
type~I extrema also implies that the dimension of the space of
solutions to a given eigenvalue is {\it even} (there are no
$SO(2)$-invariant eigenmodes in this case). We thus can arrange matters so
that $\lambda_{(2m)}=\lambda_{(2m-1)}, m=1,\dots,N$. 

There are on the other hand {\it three} zero-modes  for type~II
extrema, which I will label $\phi^{(1)},\phi^{(2)},\phi^{(3)}$,
with $\lambda_{(1)}=\lambda_{(2)}=\lambda_{(3)}=0$. 

In a semi-classical
evaluation of $\widetilde{\cal Z}[1]$ the zero-modes have to be handled with
care. The introduction of collective coordinates for the bosonic zero
modes is standard\cite{Ra87}: 
\begin{itemize}
\item[i)] The representatives in a class $[\tilde g]_I$ of type~I extrema,
are described by two collective angles $\theta,\varphi$ which (for
instance)  denote the direction of $\hat s_0$, the spin at a particular site.
\item[ii)] a particular representative  in a class $[\tilde g]_{II}$ of
type~II extrema  is specified by three collective angles
$\theta,\varphi,\psi$. $\theta,\varphi$ again give the direction of $\hat
s_0$, while $\psi\in [0,\pi]$ can be chosen to denote the direction
of $\hat s_0\times \hat s_j$, where $\hat s_j$ is a particular spin
that is {\it not} collinear to $\hat s_0$. The range of $\psi$ is
restricted to $[0,\pi]$, since $\psi\in[\pi, 2\pi]$ are
equivalent configurations (as can be seen by interchanging the
meaning of $\hat s_0$ and $\hat s_j$ in the above definitions of the
angles).  These three collective angles parameterize an $S_3$.   
\end{itemize}

In terms of the angles $\theta_i,\varphi_i$ parameterizing the
coset $SU(2)/U(1)$ as in\equ{parameterg}, the Haar-measure $dg_i$ is
proportional to
\begin{equation}
\int dg_i\rightarrow \int dh_i\int_{S_2} d\Omega_i=\int dh_i
\int_0^\pi \sin\theta_i\, d\theta_i\int_0^{2\pi} d\varphi_i\ , 
\label{measureS2}
\end{equation}
where $dh_i$ is the Haar-measure of the
$U(1)$-group. After the change of variables\equ{coeffs}  
the bosonic measure in\equ{TQFT} for sufficiently small fluctuations
$\xi^{(n)}$ near an extremal solution $\tilde g$ thus becomes
\begin{equation}
\left.\prod_{sites} dg_i\right|_{\tilde
g}=V_{{\cal H}}\prod_{n=1}^{2N} d\xi^{(n)}\ .
\label{linmeasure}
\end{equation}
For $\lambda_{(n)}\neq 0$ the fluctuation $\xi^{(n)}$ is of order
$\sqrt{\alpha}$ and the approximation\equ{linmeasure} in this case is valid
for sufficiently small $\alpha$. The coefficients of bosonic zero-modes on the
other hand are not suppressed. These fluctuations generate
(small) $SO(3)$ rotations of the extremal configuration $\tilde
g$ and are replaced by integrations over the collective
coordinates of the extremal configurations in the corresponding class
$[\tilde g]$.  The correct semi-classical measure for the
integration of bosonic fluctuations around a class of type~I extrema thus is  
\begin{equation}
\left.\prod_{sites} dg_i\right|_{[\tilde
g]_I}=V_{{\cal H}} \left(\int_{S_2} d\Omega_2\right) \prod_{n=3}^{2N}
d\xi^{(n)}\ ,
\label{measI}
\end{equation}
where $d\Omega_2=\sin\theta d\theta d\phi$ is the parameterization of
the $S_2$ in terms of the collective
coordinates. Similarly the semi-classical measure for a class of type~II
extrema is
\begin{equation}
\left.\prod_{sites} dg_i\right|_{[\tilde
g]_{II}}=V_{{\cal H}} \left(\int_{S_3} d\Omega_3\right) \prod_{n=4}^{2N}
d\xi^{(n)}\ ,
\label{measII}
\end{equation}
where $d\Omega_3=\sin\theta d\theta d\phi\sin^2\psi d\psi$ is the
parameterization of $S_3$ in terms of the collective angles.
The Jacobian for the change of basis\equ{coeffs} is a constant  and the
measure for the Grassmann-coefficients $c^{(n)},\bar c^{(n)}$ thus can
be written 
\begin{equation}
\prod_{sites} d^2c_i d^2\bar c_i=\prod_{n=1}^{2N} dc^{(n)}d\bar
c^{(n)}
\label{Gmeas}
\end{equation}

Using\equ{constZ},\equ{diagonal} and the appropriate semi-classical
measures\equ{measI},\equ{measII} and\equ{Gmeas}, the saddle point
evaluation of $\widetilde{\cal Z}[1]$ gives
\begin{eqnarray}
\widetilde{\cal Z}[1]&=&\lim_{\alpha\rightarrow 0_+} \alpha^{-N} \int
\prod_{sites\ i} dh_i d\Omega_i d^2c_i d^2\bar c_i  e^{- S_{GF}^{\rm 
eff}[1^{g},c,\bar c;\alpha]}\nonumber\\
&=&V_{{\cal H}} \left\{\sum_{[\tilde g]_I} Z_I([\tilde
g]_I)+\sum_{[\tilde g]_{II}} Z_{II}([\tilde g]_{II})\right\}\ , 
\label{semiclassical}
\end{eqnarray}
with the semi-classical weights
\begin{equation}
Z_I([\tilde g]_I)=\lim_{\alpha\rightarrow 0_+}\alpha^{-N}
\int_{S_2} d\Omega_2
\prod_{n=3}^{2N} d\xi^{(n)}\prod_{n=1}^{2N} dc^{(n)}d\bar c^{(n)} 
\exp\left\{-S_{GF}^{\rm eff}[\tilde g(\Omega_2),\alpha;
\{\xi^{(n)},c^{(n)},\bar c^{(n)}\}]\right\}\ ,
\label{ZI}
\end{equation}
and
\begin{equation}
Z_{II}([\tilde g]_{II})=\lim_{\alpha\rightarrow 0_+} \alpha^{-N} 
\int_{S_3} d\Omega_3 \prod_{n=4}^{2N} d\xi^{(n)}\prod_{n=1}^{2N}
dc^{(n)}d\bar c^{(n)} \exp\left\{-S_{GF}^{\rm eff}[\tilde g(\Omega_3),\alpha;
\{\xi^{(n)},c^{(n)},\bar c^{(n)}\}]\right\}\ ,
\label{ZII}
\end{equation}
of a class of extrema of type~I, respectively  type~II. The crucial
observation that enables us to actually compute $\widetilde{\cal
Z}[1]$ is that the weight $Z_{II}$  vanishes. It vanishes due to the
$3^{rd}$ zero-mode of  type~II  
extrema. The argument goes as  follows. In\equ{ZII} we may perform
the bosonic and fermionic integrations of all modes except the
zero-modes corresponding to $n=1,2$ or $3$. The integrals are
Gaussian and the quartic ghost interaction in\equ{diagonal} to leading
order in $\alpha$ does not contribute to these integrations. The
Grassmann integration of a pair $c^{(n)},\bar c^{(n)}$ and the
corresponding bosonic integral over $\xi^{(n)}$ for $n\neq 1,2$ or $3$
results in a factor proportional to  
\begin{equation}
\lambda_{(n)} \sqrt{\frac{\alpha}{(\lambda_{(n)})^2}}=\pm
\sqrt{\alpha} \ ,
\label{intnonzero}
\end{equation}
depending on whether $\lambda_{(n)}$ is a positive or negative eigenvalue
($\lambda^{(n)}\neq 0$ for
$n\neq 1,2,3$). We can perform $2N-3$ integrals 
in this fashion and the expression for $Z_{II}$ to leading order in
$\alpha$ (up to an irrelevant finite and $\alpha$-independent
normalization) becomes  
\begin{eqnarray} 
Z_{II}([\tilde g]_{II})&=&\lim_{\alpha\rightarrow 0_+} \pm \alpha^{-3/2} 
\int_{S_3} d\Omega_3 \int \prod_{n=1}^{3}
dc^{(n)}d\bar c^{(n)} \exp[-\alpha \sum_{klmn} R_{klmn} {\bar
c}^{(k)} {\bar c}^{(l)} {c}^{(m)} {c}^{(n)}]\nonumber\\
&=&\lim_{\alpha\rightarrow 0_+} \pm 8\pi^2 \alpha^{-1/2} 
R_{1212} \int dc^{(3)} d\bar c^{(3)}\nonumber\\
&=& 0\ .
\label{ZIIvanish}
\end{eqnarray}
The coefficient of the leading term in the loop
expansion of $Z_{II}$ vanishes due to two uncompensated Grassmann
modes. The integration over the corresponding bosonic zero-modes is
{\it finite} because $S_3$ is
compact. (This is in agreement with the Poincar\'e-Hopf theorem which
states that the contribution of such a class of extrema is
proportional to $\chi(S_3)=0$.) 
The objection that we only computed the coefficient of the term of
order $1/\sqrt{\alpha}$  and that higher
orders of the loop expansion could lead to a finite result does not
hold, because the parameter $\alpha$ in this calculation {\it is} the loop
parameter. Corrections to the above result thus are at least of order
$\sqrt{\alpha}$ and vanish in the limit $\alpha\rightarrow 0_+$.
The weight $Z_{II}$ of a single class of type~II extrema therefore
indeed vanishes.  With a finite number of spins one furthermore
expects only a finite number of such classes. In this case the total
contribution of type~II extrema to $\widetilde{\cal Z}[1]$ also
certainly vanishes. Thus the number of
classes of type~II extrema for the 1-dimensional periodic spin chain
is given by its length. Since 
none of our arguments explicitly depend on the dimensionality of the
lattice, it is safe to conclude that type~II extrema
give a vanishing contribution to $\widetilde {\cal Z}[1]$ on any finite
periodic lattice.

The semi-classical weight of a class $[\tilde g]_I$ of type~I
extrema on the other hand does {\it not} vanish. These are solutions
of\equ{extrema} where all the spins are collinear. As noted before, the
$SO(2)$-invariance of such an extremal configuration implies that for
every eigenvector $\phi^{(2m)}$, there is also an orthogonal one
$\phi^{(2m-1)}$ to the {\it same} eigenvalue. The latter is just an
$SO(2)$ rotation by $90^o$ around the common spin axis of the first.
Eigenvalues thus come in pairs. Proceeding as before and
performing the $2(N-1)$ bosonic and Grassmann integrals over
$\xi^{(n)}, c^{(n)}$ and $\bar c^{(n)}$ with $n\neq 1,2$
in\equ{ZI}, one obtains (again up to an irrelevant finite and
$\alpha$-independent overall normalization) 
\begin{eqnarray} 
Z_{I}([\tilde g]_{I})&=&\lim_{\alpha\rightarrow 0_+} \alpha^{-1} 
\int_{S_2} d\Omega_2 \int \prod_{n=1}^{2}
dc^{(n)}d\bar c^{(n)} \exp[-\alpha \sum_{klmn} R_{klmn} {\bar
c}^{(k)} {\bar c}^{(l)} {c}^{(m)} {c}^{(n)}]\nonumber\\
&=& 4\pi (4 R_{1212})=8\pi/N \ ,
\label{ZIint}
\end{eqnarray}
for the weight of any class of type~I extrema. To evaluate $R_{1212}$
in\equ{ZIint} I used that the zero-modes $\phi^{(1)}$ and $\phi^{(2)}$
correspond to global rotations of the extremal configuration and are
normalized by\equ{orthonormal}. For a collinear spin configuration
of type~I these zero-modes are readily found and the result for
$R_{1212}$ defined by\equ{R} does not depend on the (collinear)
configuration.   Note that the semi-classical 
weight $Z_I$ of each class is the same. By suitably
normalizing the Haar-measure, we can thus set $Z_I([\tilde
g]_I)=2$ for any class $[\tilde g]_I$. (I choose this normalization of
the weight in accordance with the Poincar\'e-Hopf theorem, where the
contribution of an $S_2$ manifold of extremal solutions is normalized
to $\chi(S_2)=2$.) Relative to the direction of one of the
spins, the other collinear spin can be either parallel or
anti-parallel. There are thus 
$2^{(N-1)}$ classes of extremal configurations of type~I and we
finally obtain (with the conventional normalization) 
\begin{equation}
\widetilde{\cal Z}[1]= 2^N=\chi((S_2)^N)=\chi({\cal G}/{\cal H})\neq 0\ ,
\label{nonvanishZ}
\end{equation}
in complete agreement with the Poincar\'e-Hopf theorem.

\section{Gauge fixing of the residual $U(1)$ gauge group} 
In the last three sections we have shown that the partially gauge
fixed $U(1)$-invariant LGT is normalizable and reproduces the
expectation values of gauge-invariant physical observables of the
original $SU(2)$-LGT. The lattice action\equ{effaction} is local,
invariant with respect to the abelian  
lattice gauge group ${\cal H}$ 
and preserves the space-time symmetries of the lattice. This
model could be of considerable interest in the numerical   
investigation of LGT because its structure group is abelian.
Following the procedures of\cite{Sa80} one perhaps can also derive a
corresponding {\it dual} lattice model.  

A perturbative evaluation however requires a further reduction of the
$U(1)$ structure group to a discrete one. Forcrand and
Hetrick\cite{Fo95}  presented an
elegant algorithm to {\it uniquely} and covariantly fix the gauge of an
abelian LGT by Hodge decomposition. Their procedure  solves the
problem of covariant abelian gauge fixing from a numerical point of view. The
algorithm is however non-local and I have not been able to derive the
corresponding effective gauge fixed action it generates. Recently an 
alternative solution\cite{Te98} was suggested that 
corresponds to a certain coherent superposition of Sharpe's
gauges\cite{Sh84}. To apply these gauge-fixing 
ideas to the abelian subgroup of $SU(2)$ is not entirely trivial nor 
very transparent and will not be pursued here.  It has been
argued\cite{Sh84,Bo98} that a BRST-symmetric local ``covariant''
lattice action of the link-variables that is physically equivalent to
a $U(1)$-LGT with well-defined lowest order continuum propagators does not
exist. This is in agreement with our topological 
considerations. From the topological point of view this problem is a
consequence of the fact that $\chi(U(1)/Z_n)=\chi(U(1))=0$ for any
(finite) discrete subgroup 
$Z_n\subset U(1)$. The partition function of a TLT that 
localizes on a gauge fixing surface derivable from a Morse potential
in the $U(1)$ case thus vanishes (and consequently also the partition
function of the ``gauge-fixed'' BRST-invariant model). Unfortunately
the ``linear'' covariant  gauge condition that gives well-defined
continuum propagators {\it is} the Lie-derivative of a Morse
function\cite{Zw82}.   
 
Requiring that a non-abelian gauge-fixed local lattice action leads to
well-defined propagators in the (naive) continuum limit could however
simply be too much to ask -- the continuum model is after all related to the  
continuum gauge group, which is non-compact and topologically quite
different from the compact  structure  group of the lattice. We in fact
{\it can} demand that a loop expansion of the gauge-fixed {\it
non-abelian} LGT makes sense although the naive continuum propagators
are ill-defined. One should stress in this context that a loop expansion
of lattice correlators coincides with the conventional perturbative
expansion only for vanishing (bare) coupling $g^2$. The loop expansion
is obtained by expanding the {\it full} bosonic {\it lattice} measure
in the vicinity of its maximum for small but {\it finite} bare
coupling. Additional quadratic terms of sub-leading order in $g^2$
arise in such an expansion of the effective
action from the Haar-measure as well as the ghost- (and possibly the
fermionic-) determinants. The quadratic terms from the
ghost-determinant and the Haar- measure 
generally are not transverse in a non-abelian LGT and thus lead to
well-defined {\it lattice propagators}. For sufficiently
small coupling $g^2$, the transverse (physical) part of these
lattice propagators is dominated by the naive continuum expression
whereas the longitudinal part is formally of order $1/g^2$. The loop
expansion should nevertheless result in an {\it analytic}
$g^2$-expansion of gauge  
invariant (physical) lattice correlators. The loop expansion
of unphysical correlation functions generally will not be analytic in
$g^2$.  This systematic expansion of the bosonic lattice measure 
is perhaps rather similar to the  phenomenologically successful {\it
tadpole improved} lattice perturbation theory\cite{Le93}.           

One {\it can} isolate  the classical configuration
maximizing the measure with respect to (lattice)
gauge transformations by  reducing the gauge symmetry of the abelian LGT covariantly while preserving a 
BRST-symmetry. The partition function of the corresponding  TLT simply {\it
must not} be proportional to the Euler characteristic of the $U(1)$
structure  group. One has to choose some {\it non-vanishing} topological
invariant of the group manifold -- such as the number of
connected components.  The construction of such a TLT below is based
on a nilpotent BRST-symmetry $\delta$ and proves that a local
$Z_2$-LGT is physically equivalent to the original $SU(2)$-LGT.
A numerical simulation of this $Z_2$-LGT is furthermore hardly more
complicated than the  simulation of 
the partially fixed $U(1)$-invariant model.    

The key to a further reduction of the continuous gauge symmetry of the
$U(1)$-invariant lattice theory in our case is the
Nakanishi-Lautrup field $b_i$ that was introduced by the
equivariant BRST-construction. It is a   
hermitian scalar that is {\it charged} under the $U(1)$. 
At every site it is of the form 
\begin{equation}
b_i = B_i\tau_+ + B_i^* \tau_-\ ,
\label{B}
\end{equation}
where  $B_i$ is a complex number and $B_i^*$ its complex
conjugate. Parameterizing $h_i=e^{2 i\varphi_i\tau_0}\in U(1)$, with
$\varphi\in[0,2\pi)$, one observes that $B_i$ transforms as
\begin{equation}
B_i^h=e^{2i\varphi_i} B_i\ ,
\label{transformB}
\end{equation}
under the residual $U(1)$.  
We thus have the option to fix the phase of the Nakanishi-Lautrup field and
thereby reduce the gauge invariance of the model to the discrete gauge
group $Z_2\subset U(1)$ (we can't do better, since the
Nakanishi-Lautrup field  is oblivious to 
$Z_2$ gauge transformations of the lattice configuration). 
We can for instance require that $B$ is a real
and {\it positive} field. The corresponding measure in the gauge fixed
functional integral becomes
\begin{equation}
\int d^2 b_j\quad {\stackrel{{U(1)\rightarrow
Z_2}}{-\!\!\!-\!\!\!-\!\!\!-\!\!\!-\!\!\!\!\longrightarrow} }\quad
\int_0^\infty B_j dB_j\ . 
\label{chmeasure}
\end{equation}      
Note that the integration is  over {\it positive} real variables $B_j$ 
only. (An unconstrained integration over all real values of $B_j$ would
lead to a vanishing partition 
function of the corresponding TLT which can then be shown to be
proportional to the Euler character of $S_1$.) 

We can  perform the integration of the Nakanishi-Lautrup field $b_i$
also in the $Z_2$-LGT. Due to\equ{direction} the anti-hermitian field
$F_j[U]$ at each site is given by a complex number $f_j[U]$ and may be
written
\begin{equation}
F_j[U]=\tau_+ f_j[U] -\tau_- f_j^*[U]\ .
\label{f}
\end{equation}
Using\equ{chmeasure} and\equ{onshell} the $B$-dependent bosonic part
of the partition function for the $Z_2$ LGT is local and
proportional to 
\begin{equation}
\prod_{sites}\int_0^\infty B_i dB_i e^{-2 i B_i{\rm Im} f_i[U] -
\alpha B_i^2} \propto \prod_{sites} {\cal P}(\frac{{\rm Im}
f_i[U]}{\sqrt{\alpha}})\ , 
\label{Bint}
\end{equation}
with the complex weight function ${\cal P}$,
\begin{equation}
{\cal P}(x)=1- x e^{-x^2}\left( i\sqrt{\pi} +\int_{-x}^{x} e^{t^2}
dt\right)\ ,
\label{P}
\end{equation} 
related to the error function on the imaginary axis. 
Note that ${\cal
P}(-x)={\cal P}^*(x)$ and that $f_i[U]=-f_i[U^\dagger]$, i.e. $f_i$
changes sign when the direction of all the links is reversed. The expectation
value of  hermitian  observables of the link variables is thus real. 
From\equ{P} we observe that the local measure ${\cal
P}(x\sim\infty)\sim 1/(2 x^2)$ does not vanish exponentially for large
values of $x={\rm Im} f_i[U]/\sqrt{\alpha}$. $|{\cal P}(x)|$ however
decreases monotonically and peaks at $x=0$. For sufficiently large
lattices ${\cal P}(x)$ can be approximated by
${\cal P}(x)\sim \exp[-i\sqrt{\pi} x -(2-\pi/2) x^2]$ up to terms that
are irrelevant in the critical limit of the model. Note that the total
phase of the measure in this limit is proportional to $\sum_i {\rm Im}
f_i[U]$ and vanishes on a  periodic lattice. Expanding near the
trivial configuration $U=1$, it is readily 
seen  that ${\rm Im} f_i[U\sim 1]$ is {\it not} linear in all
longitudinal fluctuations. In agreement with the  previous discussion,
the modification of the lattice action by a
term proportional to$ ({\rm Im} f_i[U])^2$  does not lead to   
well defined (naive) continuum propagators.

To exhibit the
BRST-structure of this $U(1)$ gauge fixing and 
relate it to a TLT, we note that it can be obtained  by inserting 
\begin{equation}
{\cal Z}_{U(1)}[B]=\prod_{sites} Z[B_i]\ ,
\label{factorizeZ}
\end{equation}
in the functional integral. Here the local TLT ``partition function''
at each site is simply the integral 
\begin{equation}
Z[x]= \oint_{|z|=1} \!\!\! z^*dz
|xz+x^*z^*| \delta(i(xz-x^*z^*))\ . 
\label{TQI}
\end{equation} 
The integration of the $U(1)$-group element $h_i\in U(1)$  in\equ{TQI}
is here written as a contour integral of $z$ along the unit circle in
the complex plane.
\equ{transformB} shows that one in principle has to integrate twice
over the unit circle. This however just introduces an irrelevant factor of
$2$ in \equ{TQI}.

If $Z[x]$ is a non-vanishing constant one can insert\equ{factorizeZ} in
the functional integral and change 
variables to factorize the group volume $V_{\cal H}$ and arrive at the
effective measure\equ{chmeasure} of the gauge-fixed model.
Performing the
integrations in\equ{TQI} one
explicitly finds  that $Z[x]$ does not vanish and furthermore does not
depend on $x$. The integral $Z[x]$ is thus
a topological invariant of the circle $S_1$. The corresponding
topological model with local ``action'' can be 
explicitly constructed  by exponentiating all the factors
in\equ{TQI}. Introducing real bosonic variables $u,v$ and Grassmann variables
$\eta,\bar\eta,\nu,\bar\nu$ one can rewrite $Z[x]$ as the integral
\begin{equation}
Z[x]=\oint_{|z|=1}\!\!\! z^*dz \int\!\!\int_{-\infty}^{\infty} du dv d\eta
d\bar\eta d\nu d\bar \nu e^{-S[xz,u,v,\eta,\bar\eta,\nu,\bar\nu]}\ ,
\label{topint}
\end{equation}
with the local ``action'',
\begin{equation}
S[a,u,v,\eta,\bar\eta,\nu,\bar\nu]= u(a-a^*)+
(a+a^*)(v^2(a+a^*)+\eta\bar\eta +\nu\bar\nu)\ .
\label{acU1}
\end{equation}
Performing the Grassmann integrals over $\bar\eta,\eta,\bar\nu,\nu$
and the ordinary Gaussian integral over $v$ in\equ{topint} gives $|a+a^*|$.
The integration over $u$ leads to the constraint $a=a^*$. To show
that\equ{topint} with action\equ{acU1} is a topological integral
of Witten type, we  verify that $S$ is exact with respect to a
nilpotent symmetry $\delta$ defined on the (local) variables as 
\begin{eqnarray}
\delta x&=&0\nonumber\\
\delta z&=&z\eta,\  \delta z^*=-z^*\eta ,\ \delta \eta=0\nonumber\\
\delta a&=&\delta x z=a\eta ,\ \delta a^*=-a^*\eta\nonumber\\
\delta\bar\eta&=&u +\nu\bar\nu +v^2(a-a^*) ,\ \delta u=-\nu\bar\nu \eta
- v^2 (a+a^*)(\nu+\eta) \nonumber\\
\delta\bar\nu &=&\bar\nu (\nu-\eta)-2 v^2 a\ ,\ \delta v=-v\nu/2,\
\delta\nu=0\ . 
\label{delta}
\end{eqnarray}
Using the algebra\equ{delta} it is straightforward to show that $\delta$ is
nilpotent, $\delta^2=0$. Note that $z z^*=1$ and (consequently) $a
a^*$ are invariants. I obtained\equ{delta} by demanding $\delta
x=0$ and $\delta 
z=z\eta$, i.e. by {\it ghostifying} the $U(1)$ transformation. Together
with the nil-potency of $\delta$ these assumptions imply the first
three lines of relations in\equ{delta}. The remainder of the algebra
was found by demanding that the action\equ{acU1} is $\delta$-closed. 
I unfortunately can offer no further insight for the
construction of the algebra \equ{delta} which appears to be far too
involved for the problem it solves. At the end of the day one
obtains that the action\equ{acU1} is the $\delta$-exact expression
\begin{equation}
S[a,u,v,\eta,\bar\eta,\nu,\bar\nu]=\delta (\bar\eta (a-a^*) - 2\bar\nu
a^*)\ .
\label{dexact}
\end{equation}
This shows that\equ{topint} can be interpreted as a topological
invariant of Witten type. This invariant does not vanish and is
{\it  not} proportional to the Euler character of $U(1)$. The only 
independent topological invariant of a circle is the number of its connected
components (the lowest Betty number $b_0(S_1)=1=b_1(S_1)$). This
topological characteristic, like the Euler character of a manifold,  is
multiplicative, i.e. $b_0(M_1\times 
M_2)=b_0(M_1) b_0(M_2)$ for any two compact manifolds $M_1$ and
$M_2$. We can evidently choose to normalize\equ{TQI} so that
$Z[x]=b_0(S_1)=1$ and then interpret the partition function ${\cal
Z}_{U(1)}[B]$ of\equ{factorizeZ} as the number of connected components
of the $U(1)$ gauge group of the lattice. 
The TLT construction supports the claim\cite{Ba98} that gauge fixing is 
equivalent to the construction of a TQFT on the gauge group whose
partition function does {\it not} vanish. 

\section{The auxiliary field $\rho $}   
In Appendix~A the Grassmannian fields are integrated 
in favor of a non-local measure\equ{bosonint} for the gauge
fixed LGT. It is thus in principle possible to numerically simulate
the $Z_2$-LGT. We have shown that such a
simulation would reveal nothing new for {\it gauge
invariant} physical observables. The reason for constructing a physically  equivalent covariant LGT with a much smaller 
invariance group was to gain a better {\it analytical} understanding of the
model in the critical
limit. The best we could  do was 
a reduction of the continuous $SU(2)$ gauge symmetry of the original
LGT to a discrete $Z_2$-structure group. The natural question to ask is
whether this discrete gauge group is spontaneously broken. Although
there is no gauge invariant physical order parameter, whether or not
this symmetry is broken could shed some light on the dynamics of the
model.   

Here I will however only discuss the role of the auxiliary scalar
field $\rho_i$ introduced to linearize the quartic ghost
interaction.  I will show that the bosonic measure is maximal at a
non-trivial (constant) $\rho_i$. This is crucial  for a loop
expansion of physical observables of the lattice model. I first show
that the effective measure for the field $\rho_i$ is non-trivial and 
gauge invariant. Consider the weight of\equ{gfweight}
as a function of the link configuration $U$ and the auxiliary
variable $\rho$ integrated over the gauge group ${\cal G}$:
\begin{equation}
{\cal Q}[U,\rho; \alpha]=\int \prod_{sites} dg_i {\cal W}[U^g,\rho;\alpha]
\label{physrho}
\end{equation}
By construction the observable ${\cal Q}$ is gauge invariant 
\begin{equation}
{\cal Q}[U^g,\rho;\alpha]={\cal Q}[U,\rho;\alpha]\ ,
\label{invQ}
\end{equation}  
for any configuration $\rho$. We furthermore know from the previous
sections that on any finite lattice, 
\begin{equation}
{\cal N}(\alpha)= \int \prod_{sites} d\rho_i {\cal Q}[U,\rho;\alpha] ,
\label{constQ}
\end{equation} 
is a non-vanishing (finite) normalization constant that does not depend
on the link configuration $U$. The two results\equ{invQ} and
\equ{constQ} imply that one can define a normalizable measure $W[\rho]$
for the scalar field $\rho$ 
\begin{equation} 
W[\rho;\alpha, g^2]=\vev{{\cal Q}[U,\rho;\alpha]}_{inv.} \ . 
\label{W}
\end{equation}
Here the expectation value on the RHS is with the original
SU(2)-invariant measure. \equ{constQ} implies
that $W[\rho]$ is normalizable and does not vanish identically. The
measure\equ{W} for the configurations $\rho$ is therefore  non-trivial.
Upon changing variables $U\rightarrow U^g$ in the
gauge-invariant RHS of\equ{W} we can decouple the
integration over the gauge group ${\cal G}$ and equivalently write 
\begin{equation}
W[\rho;\alpha, g^2]=\vev{1}_\rho\ ,
\label{vevh}
\end{equation}
where $\vev{{\cal O}}_\rho$, given by 
\begin{equation}
\vev{{\cal O}}_\rho =\int \prod_{links} dU_{ij} {\cal O}[U,\rho]
{\cal W}[U,\rho;\alpha] e^{-S_{\rm inv.}[U]}\ ,  
\label{intrho}
\end{equation} 
is the expectation value of a {\it gauge invariant} function $
{\cal O}[U,\rho]={\cal O}[U^g,\rho]$ for a given configuration
$\rho$ of the {\it gauge-fixed} model. Due to\equ{physrho}
computing\equ{intrho} for gauge invariant 
observables ${\cal O}$ is entirely equivalent to evaluating the
gauge-invariant correlator 
\begin{equation}
\vev{{\cal O}}_\rho =\vev{{\cal Q}[U,\rho;\alpha]\  {\cal O}}_{inv.}\ ,
\label{equiv}
\end{equation}    
with the  SU(2)-invariant measure of the $SU(2)$-LGT. 

A perturbative evaluation of the $Z_2$-LGT  
should at least  retain those configurations $\bar\rho$ for which
$\vev{{\cal O}}_{\bar\rho}$ (and in particular $\vev{1}_{\bar\rho}$)
{\it do not} vanish in the limit $g^2\rightarrow 0$. I argue that
these are  configurations in the vicinity of non-trivial constant
configurations $\bar\rho_i=const.$ only.

In the limit $g^2\rightarrow 0$, the gauge invariant action $S_{\rm inv.}[U]$
on a {\it finite} lattice constrains the configuration space $U$ to
the subset  of pure gauge configurations $U_{ij}=g_i g^\dagger_j$. As
we have seen in section~IV there are at least two zero modes of the
quadratic form  in\equ{effSGF} in this case: they correspond to {\it
global} rotations of the gauge spins $\hat s_i$. These zero-modes were
shown to be absorbed by the quartic ghost interaction of\equ{effSGF}. In the
linearized version\equ{linSGF} of the model these
zero-modes couple to the auxiliary field $\rho$ only. The determinant
in\equ{gfweight} and consequently\equ{intrho} for a finite lattice 
thus vanish in the limit
$g^2\rightarrow 0$ for configurations $\rho$ that satisfy
\begin{equation} 
\sum_i \rho_i\Tr [\tau_0, \phi_i^{(1)\dagger}] \phi_i^{(2)} =0 \ ,    
\label{cond}
\end{equation}
where $\phi_i^{(1)},\phi_i^{(2)}$ are the two global zero modes  of a
pure gauge configuration. Since the 
{\it global} zero-modes rotate all the spins equally, the vector $v_i$
\begin{equation}  
\Tr [\tau_0, \phi_i^{(1)\dagger}] \phi_i^{(2)}=v_i= const.\ ,
\label{global}
\end{equation}
has constant entries irrespective of the pure gauge
configuration being considered. \equ{cond} implies that
$\vev{1}_\rho\rightarrow 0$ in the 
limit $g^2\rightarrow 0$ on any {\it finite} lattice if the
configuration $\rho$ is orthogonal to $v_i$, i.e. has no constant
component. Note that this is true for any value of $\alpha\neq 0$.

The argument above is however  true only in the {\it wrong}
limit where one takes  $g^2\rightarrow 0$ {\it before} considering the
thermodynamic limit $N\rightarrow\infty$. In the thermodynamic limit
one can only say that the relevant configurations in the limit
$g^2\sim 0$ are (in a statistical sense)
in the vicinity of pure gauge 
configurations. One nevertheless would expect that  configurations
$\rho$ with large contributions to $\vev{1}_\rho$ are also in some
sense close to nontrivial constant ones. More precisely, the argument
above and the considerations of section~III indicate that a
loop expansion of the gauge fixed $Z_2$-model on a {\it finite}
lattice in the vicinity of a (particular) pure gauge configuration
is sensible only for $\rho_i=\bar\rho_i=const.\neq 0$.  We
otherwise would expand about a configuration that has vanishing
weight.

The covariant loop expansion of the $Z_2$-model can in fact be examined in
more detail and also  gives some insight into the critical limit of
the model on an infinite lattice.
Consistency requires that the value of $\bar\rho$ be determined 
order by order of the loop expansion by
\begin{equation}
\vev{\rho_i}=\bar\rho\ ,
\label{gap}
\end{equation}
with $\vev{\rho_i}$ given by\equ{bosonint}. At the
``tree''-level of the loop expansion for the {\it bosonic}
measure\equ{gap}  implies that the unique maximum of the 
measure is at $\bar\rho^{tree}$ which is the solution of
\begin{equation} 
\left.\frac{\partial}{\partial \bar\rho} {\cal W}[U^{tree},\bar\rho;
\alpha]\right|_{\bar\rho^{tree}} =0\ ,  
\label{extreme}
\end{equation}
for a pure gauge configuration $U^{tree}$ that 
satisfies the gauge condition,
\begin{equation}
{\rm Im} f_i[U^{tree}]=0\ .
\label{constr}
\end{equation}
As far as the perturbative expansion 
of gauge invariant observables is concerned, we may choose any
{\it one} of the $2^N$ (gauge equivalent) configurations $U^{tree}$ that
contribute to the partition function. These were obtained in section~IV
and correspond to collinear gauge spins. We may then solve
\equ{extreme} to obtain the appropriate value of $\bar\rho^{tree}$. In
fact, $\bar\rho^{tree}$ is the same for any one of the $2^N$ 
discrete ``vacua'' due to the gauge invariance\equ{W} of $W[\rho]$. 

The simplest (perturbative) vacuum configuration for the links, and
the only one that leads to a covariant perturbative expansion, is
$U^{tree}_{ij}=1$.  With periodic boundary conditions for the lattice,
$\det{\cal M}[u=1,v=0,\rho]$ is readily calculated for constant
$\rho_i=\bar\rho$. One can diagonalize ${\cal
M}[1,0,\bar\rho]$ in the basis of eigenvectors $X_i(\vec n)$ of the
hermitian matrix $A[1]$ given by\equ{defmat}   
\begin{equation}
A[1]\cdot X(\vec n)=\Delta_{{\vec n}} X(\vec n)      
\label{laplace}
\end{equation}
where 
\begin{equation}
\Delta_{{\vec n}}=-4\sum_{\mu=1}^D \sin^2(\pi n_\mu/L)\ ,\
n_\mu=1,\dots L
\label{Laplaceeigen}
\end{equation}
are the eigenvalues of the Laplace-operator of a $D$-dimensional
hyper-cubic lattice with $L^D=N$ sites. The eigenvalues of ${\cal
M}[1,0,\bar\rho]$ are 
\begin{equation}
\lambda^{(\vec n)}_\pm=\Delta_{{\vec n}} \pm i\bar\rho\ ,
\label{eigen}
\end{equation}    
and $\det{\cal M}$ for the vacuum configuration therefore is
\begin{equation}
\det{\cal M}[1,0,\bar\rho]=\prod_{\vec n} \left( \Delta_{{\vec n}}^2
+ \bar\rho^2\right)\ .
\label{det0}
\end{equation}
As expected, the determinant\equ{det0} vanishes for
$\bar\rho\rightarrow 0$ like $\bar\rho^2$ on any finite lattice. 
The determinant\equ{det0} is furthermore a monotonically increasing
positive function of $\bar\rho^2$. For $\bar\rho^2\gg D$, the
determinant behaves as ${\cal M}[1,0,\bar\rho\gg D]\sim
\bar\rho^{2N}=e^{N\ln\bar\rho^2}$. Comparing with the monotonically
decreasing exponent
$e^{-N\bar\rho^2/(4\alpha)}$ in ${\cal W}$, we 
see that the maximum is {\it unique}  and of order $\bar\rho^2\sim
4\alpha$ for large $\alpha$. For a perturbative expansion we 
are however interested in the value for the maximum at (arbitrary) small
$\alpha\sim 0$ for which gauge fluctuations are suppressed.
In this limit we expect $\bar\rho^2\sim 0$ also. 
Using\equ{det0},\equ{gfweight} and the definition\equ{W},  
the unique value $\bar\rho(\alpha)$ where the weight
$W[\bar\rho(\alpha);\alpha,g^2\sim 0]$ is maximal to lowest order in
the loop expansion is the solution of 
\begin{equation}
\sum_{\vec n} \frac{1}{ \Delta_{{\vec n}}^2 +
\bar\rho^2}=\frac{N}{4\alpha}\ .
\label{gapeq}
\end{equation}
For a {\it finite} lattice this gap equation would have to be solved
numerically. In the thermodynamic limit, the summations in\equ{gapeq} 
can be performed. In this limit\equ{gapeq} on a
periodic $D$-dimensional lattice becomes
\begin{equation}
\int_0^\infty dx\frac{\sin(x\bar\rho)}{\bar\rho}\left( e^{-2 x}
I_0(2 x)\right)^D=\frac{1}{4\alpha}\ ,
\label{thgap}
\end{equation}
where $I_0(x)$ is the Bessel function of zeroth order at imaginary
arguments.  \equ{thgap} is obtained by exponentiating the summand
in\equ{gapeq} and using the identity
\begin{equation}
\lim_{L\rightarrow\infty} \frac{1}{L}\sum_{n=1}^L e^{-2 x \sin^2(\pi
n/L)} =\frac{e^{-x}}{2\pi}\int_0^{2\pi} e^{x\cos\phi} d\phi
=e^{-x} I_0(x)\ .
\label{trick}
\end{equation}
The asymptotic behavior of $I_0(x\sim\infty)\sim e^x/\sqrt{2\pi x}$
shows that the integral in\equ{thgap} converges for $\bar\rho\neq 0$
in any dimension. In $D<4$ dimensions the integral behaves like
$\bar\rho^{(D-4)/2}$ for $\bar\rho\sim 0$. At the upper critical
dimension $D=4$ its behavior is logarithmic. For sufficiently small
$\alpha$, \equ{thgap} in $D=4$ is well approximated by
\begin{equation}
\ln \frac{\bar\rho^2}{\kappa^4} = -\frac{2 \pi^2}{\alpha}
+O(\bar\rho)\ ,
\label{detrho}
\end{equation}
where the constant $\kappa$ is given by 
\begin{eqnarray}
\ln \frac{\kappa^2}{4\pi}&=& 1-\gamma_E +4\pi^2 \int_0^\infty dx
x\left[\left(I_0(x) e^{-x}\right)^4 -\frac{1}{1+4 \pi^2
x^2}\right]\nonumber\\
&=& 2.26098\dots 
\label{kappa}
\end{eqnarray}  
For $\alpha\sim 0$, \equ{detrho} determines the optimal $\bar\rho$ of the
corresponding vacuum configuration $(U=1,\bar\rho)$ for 
the perturbative expansion in $D=4$ dimensions. Note that the relation
\equ{detrho} is not affected by  physical fermions
(i.e. quarks). The corresponding fermionic determinant is gauge
invariant, does not depend on the auxiliary field $\rho$ and does not
vanish at the pure gauge configuration 
$U=1$. This contribution to lowest order in the loop expansion can
thus be absorbed in the normalization of $W$. The  critical limit
$g^2,\alpha\rightarrow 0$, of the LGT can be interpreted as a quantum
field theory by assigning  a spacing $a$ to the lattice and defining
continuum fields. The continuum field  $\tilde 
\rho(x)$ has canonical dimension~2 and is related to $\rho$ by
$\rho_i=a^2\tilde\rho(x_i)$.
In the critical  limit \equ{detrho} is the tree-level statement 
\begin{equation}
\vev{\tilde\rho(x)}=(\kappa^2/a^2) e^{-\frac{\pi^2}{\alpha}}\ ,
\label{exprho}    
\end{equation}
for the continuum field $\tilde\rho(x)$. Due to\equ{W},
$\vev{\tilde\rho}$ is a physical gauge invariant quantity and we
could regard the LHS of\equ{exprho} as a constant physical
scale of the model. In this case \equ{exprho} is the expression for
dimensional transmutation of the gauge parameter $\alpha$ to lowest
order. On the other hand the physical asymptotic scale parameter
$\Lambda_L$ of the lattice to this order in the loop expansion is
related to the coupling $g^2$ and the lattice spacing $a$ by 
\begin{equation}
a^2 \Lambda_L^2= e^{-\frac{24\pi^2}{(11-n_f) g^2} +O(\ln g^2)}\ .
\label{Lamlattice}
\end{equation}
$\Lambda_L$ is
a finite physical scale  in the critical limit of the
SU(2)-LGT with $n_f$ fermionic flavors. \equ{exprho} singles out  
a {\it particular} 
gauge $\bar\alpha(g)=g^2 (11-n_f)/24 + O(g^4)$ in which $\vev{\tilde\rho(x)}$ scales
like a physical quantity in the critical limit, that is
\begin{equation}
\vev{\tilde\rho}=\kappa^2\Lambda_L^2 \text{ for
}\alpha=\bar\alpha=g^2(11-n_f)/24 + O(g^4)
\label{scaling}
\end{equation}
In this {\it critical} gauge the loop  expansion of gauge invariant
correlators automatically produces power corrections (in $\bar\rho\neq 0$)
that scale correctly in the critical limit. 
Note that the power corrections of the 
loop expansion vanish exponentially compared to $\Lambda_L$ in the
critical limit for gauges $\lim_{g\rightarrow
0}\alpha/g^2<(11-n_f)/24$ and dominate the correlations 
in gauges $\lim_{g\rightarrow 0}\alpha/g^2>(11-n_f)/24$ -- a sign that
the asymptotic expansion does not make much 
sense in such gauges. We know on the other hand that {\it physical} power
corrections do arise in the full theory. In critical gauges with 
$\lim_{g\rightarrow 0}\alpha/g^2=(11-n_f)/24$ they also arise in
the loop expansion of the gauge-fixed model. It is justified
to call these gauges {\it critical} because they delineate the domain
of validity of the loop expansion. 

To check the assertion that $\bar\rho$ scales like a physical quantity
in critical covariant gauges, one should evaluate the anomalous
dimension of $\tilde\rho$. From the foregoing one expects this anomalous
dimension to vanish to leading order in the gauge  $\bar\alpha(g)=g^2
(11-n_f)/24+O(g^4)$. It is then possible to adjust the critical value
$\bar\alpha(g)$ order by order in the loop expansion so that the
anomalous dimension of $\tilde\rho$ vanishes to all orders. 
I only wish to stress here that the existence of such 
a critical gauge is a direct consequence of the fact that the ghost
determinant $M[U=1,\rho]$ {\it vanishes} for $\rho=0$. If it where {\it finite}
at $\rho=0$, we would have been justified to expand 
perturbatively around  $\bar\rho=0$ for sufficiently small values of
$\alpha$. In actual fact \equ{thgap} has
{\it no solution} for $\alpha\rightarrow 0$ in $D>4$ dimensions: in
the thermodynamic limit the weight $W[\rho,\alpha]$ peaks at $\bar\rho=0$
for $\alpha$ less than some critical value in $D>4$ dimensions. In $D\leq 4$,
$\bar\rho=0$ is however only approached as $\alpha\rightarrow 0$ and the
weight $W$ is  maximal at a nontrivial value of $\bar\rho$ for {\it
any} non-vanishing  $\alpha$. 

\section{Summary and Comments}
In the foregoing we constructed a LGT with a
discrete structure group $Z_2$ that is physically equivalent to the
standard $SU(2)$-LGT.  The $Z_2$-model possesses all
of the space-time symmetries of the original LGT. The reduction of
the gauge group was shown to be equivalent to the formulation of a TLT
on a coset- respectively group- manifold. Care was taken to
ensure that the partition function of the TLT's (and consequently the
partially gauge-fixed LGT's) are normalizable. 
On a lattice, using Morse theory to construct  a TLT whose partition
function is proportional to the Euler characteristic of a compact
manifold is a mathematically rigorous procedure. We saw that this method by
itself does not suffice to fix the gauge completely because the Euler
character of the lattice gauge group ${\cal G}$ vanishes. To partially
fix the original $SU(2)$-gauge symmetry to a discrete
$Z_2$ gauge symmetry we proceeded in two steps.

The gauge invariance was first reduced to the Abelian
$U(1)$ gauge group using an equivariant BRST-construction. We showed 
that this procedure is  equivalent to the formulation of a
TLT on the coset space ${\cal G}/{\cal H}= [SU(2)/U(1)]^N=S_2^N$ and 
explicitly proved that the partition function of this 
TLT is proportional to the Euler characteristic of the coset
manifold. Since this Euler number does not vanish, the TLT is
normalizable and the partially gauge fixed $U(1)$-LGT is physically
equivalent to one with non-abelian structure group
$SU(2)$. Although we here only considered  an $SU(2)$-LGT, the
procedure can be generalized to fix the gauge of an $SU(n)$-LGT
to the maximal abelian subgroup ${\cal H}=(U(1)^{n-1})^N$. This
follows by induction from 
$SU(n+1)/(SU(n)\times U(1))\simeq S_{2n+1}/S_1\simeq CP_n$   and the
fact that the Euler character $\chi(CP_n)=n+1$ does not
vanish\footnote{I would like to thank A.~Rozenberg for this
remark.}.  In the case I considered, the  
action\equ{effaction} of the  $U(1)$-LGT is local but
depends on Grassmannian ghosts and includes a 4-ghost interaction. To
my knowledge it is the first example of a (partially) gauge-fixed
lattice model with an (equivariant) BRST-symmetry that is proven to be
physically equivalent to the original $SU(2)$-LGT also
non-perturbatively. It could be
considered the first concrete realization of non-abelian BRST-symmetry in a
non-perturbative setting. The construction of the corresponding TLT
and the proof in section~IV show   how the
Gribov-ambiguity  associated with the covariant gauge fixing is 
circumvented: although {\it there are} many Gribov copies (and  even
whole manifolds of them) associated 
with any orbit, they conspire to give a topological invariant (in
our example the Euler character $\chi(S_2^N)=2^N$) that
does not depend on the orbit within a connected sector. Since the
orbit space of a LGT is connected, the existence of Gribov copies in
covariant  gauges does {\it not} invalidate the gauge-fixing procedure
if the topological invariant the TLT computes does not vanish. This is
in contradistinction to conventional
Dirac-quantization of first class constraints\cite{Di50}, which in
principle is valid only if the solution to the gauge condition is {\it
unique}. The formulation of gauge-fixing as a topological model on the
moduli-space of the gauge theory perhaps also clarifies the
dispute\cite{Sh84,Te98,Bo98}
concerning the non-perturbative validity of covariantly gauge fixed
models with BRST-symmetry. I believe this procedure in general permits
one to handle  Gribov ambiguities.

The $U(1)$-invariant lattice model was subsequently  reduced
to one with a $Z_2$-structure group by using the  Nakanishi-Lautrup
field of the previous partial
gauge fixing.  This $U(1)$-gauge fixing is
entirely local and the constraints can be  {\it solved}
explicitly. The gauge fixing can again be related to a corresponding
local TLT. The  
partition function of this TLT is however proportional to the number of
connected components of the $U(1)$-gauge group rather than its
Euler character (which vanishes).  
The gauge fixing is  {\it not linear} and  naive continuum
propagators do not exist. I argued  that this 
should not prevent one from considering the {\it loop}
expansion of the $Z_2$-LGT since the lattice propagators 
{\it do exist} for any finite value of the coupling.   

The maxima of the measure of the $Z_2$-LGT are isolated and a
loop  expansion of gauge invariant observables in the vicinity of  
{\it any} gauge-equivalent vacuum configuration of this
model is feasible.  The price one pays is the
considerably more complicated 
non-local and generally complex measure of the resulting bosonic $Z_2$-LGT
after integration of the Grassmannian variables. 
This bosonic partition function depends  on
the link variables $U_{ij}$ {\it as well as} a local {\it gauge invariant}
scalar field $\rho_i$ and on {\it two} coupling constants
$g^2$ and $\alpha$. The former is inherited from the original
$SU(2)$-LGT whereas the latter was introduced by the gauge fixing. 

The expectation values of physical gauge invariant observables of the
original LGT
(gauge invariant functions of the link variables only) do not
depend on $\alpha$ by construction. The expectation value of gauge
invariant functions of the auxiliary field $\rho_i$
and especially $\vev{\rho_i}$ however generally {\it do} depend on $\alpha$ as
well as $\Lambda_L$.  We
found that the  maximum of the bosonic measure for  $\alpha\neq 0$ occurs at
$\rho_i =\bar\rho(\alpha)\neq 0$ and derived the {\it gap
equation}\equ{gapeq} relating $\bar\rho$ to $\alpha$ in lowest order. 
In the thermodynamic limit of a 4-dimensional lattice  the relation is
given by \equ{detrho} in  the limit of very small $\alpha\sim 0$. The
most interesting result of this analysis is that  
the expectation value $\vev{\tilde\rho(x)}$ of the corresponding
continuum field $\rho_i=a^2 \tilde\rho(x_i)$ is proportional to the
asymptotic scale parameter $\Lambda_L^2$ in a
particular {\it critical} gauge $\bar\alpha(g)= g^2(11-n_f)/24
+O(g^4)$.
{\it Non-perturbative} power corrections to physical observables
proportional to $\Lambda_L^{2k}/p^{2k}$ (with $k\geq 2$)  appear {\it
computable} in this critical 
gauge. In effect this would imply that the non-perturbative  expectation values
of Wilson's Operator-Product-Expansion for the asymptotic behavior of
physical correlators are {\it part of} the asymptotic loop
expansion in the {\it critical} gauge. Although a direct (numerical)
evaluation of physical correlations was shown to be independent of
the gauge parameter, I argued that the {\it accuracy} of the
asymptotic perturbative expansion may, and generally  {\it does}, 
depend on the gauge.  The analysis of the $SU(2)$ LGT tends to support
the conjecture that power corrections are accessible by the
loop-expansion in certain covariant gauges. A similar mechanism was
previously observed in the continuum theory\cite{Sc98}.  In this case the expectation value of a scalar moduli-parameter
also was related\cite{Sc98b} to the scale anomaly of the model. 

If power corrections to physical correlators are indeed 
computable in critical covariant gauges, the loop expansion 
in conjunction with dispersion relations could be a
powerful tool to obtain  information on the spectrum of the
model. The phenomenological success of QCD sum-rules\cite{Na89} suggests
that it might be worth pursuing this possibility.

Apart from these speculations, the topological approach to gauge fixing
of a LGT has shown that
\begin{itemize}
\item[i)] gauge fixing of a LGT is equivalent to the construction of a
certain TLT of Witten type;
\item[ii)] the gauge-fixed lattice model is
normalizable only if the topological invariant computed by the
partition function of the associated TLT does not vanish; 
\item[iii)] the BRST-symmetry of the gauge fixed LGT is inherited from
the associated TLT and is also realized non-perturbatively;
\item[iv)] covariant and BRST-invariant gauge fixing of a LGT is
possible and the Gribov ambiguity of these  gauges can be 
controlled; 
\item[v)] quartic ghost interactions arise naturally in the
non-abelian case due to residual global invariances and are perhaps
unavoidable  in covariant gauges. 
\end{itemize}
 At present this approach 
appears to be the only systematic  method that
guarantees that the gauge-fixed model is {\it covariant, local and
physically equivalent}  to the original non-abelian  gauge invariant
theory also non-perturbatively.
 
\begin{center}
ACKNOWLEDGMENTS
\end{center}
I would like to thank D.~Zwanziger, L.~Baulieu and L.~Spruch for
their invaluable support and A.~Starinets and A.~Rozenberg for endless
but valuable discussions.  

\appendix
\section{Some calculations specific to an SU(2)-LGT}
The link matrices $U_{ij}\in SU(2)$ of an SU(2)-LGT,
\begin{equation}
U_{ij}=u_{ij}(1/2 +\tau_0) +u^*_{ij}(1/2-\tau_0) +v_{ij}\tau_+
-v^*_{ij}\tau_-\ , 
\label{paramU}
\end{equation}
can be parameterized by two complex numbers $u_{ij}$ and $v_{ij}$ that
satisfy the constraint 
\begin{equation}
|u_{ij}|^2 + |v_{ij}|^2=1\ .
\label{constraint}
\end{equation}
This parameterization facilitates some calculations in the
SU(2)-LGT\cite{Cr83}. Below I give  expressions for some of
the quantities of the main text in terms of the $u_{ij}$'s and
$v_{ij}$'s. 

The Morse-potential\equ{defV} can be written,
\begin{equation}
V[u,v]=\sum_{links} |v_{ij}|^2=\sum_{links} (1-|u_{ij}|^2)\ .
\label{morseuv}
\end{equation}
U(1) gauge transformations change the phases of
$u_{ij}$ and $v_{ij}$ but not their lengths. An
infinitesimal transformation $g_i\sim 1\in SU(2)/U(1)$ is of the form  
\begin{equation}
g_i=1+\theta_i\tau_+ -\theta^*_i\tau_-\ ,
\label{inf}
\end{equation}
where the $\theta_i$ are infinitesimal complex
numbers. To first order, the parameters $u_{ij}$ and $v_{ij}$ of a
link change by
\begin{eqnarray}
\Delta u_{ij}&=&v_{ij}\theta^*_j -\theta_i v^*_{ij}\nonumber\\
\Delta v_{ij}&=&\theta_i u^*_{ij}- u_{ij}\theta_j\ .
\label{varuv}
\end{eqnarray}
The constraint\equ{constraint} to this order is invariant under the
transformation\equ{varuv}. {F}rom\equ{varuv} we obtain that the
Morse-function\equ{morseuv} changes by
\begin{eqnarray}
\Delta V[u,v]&=& \sum_{links} \theta_i v^*_{ij} u^*_{ij}
-v^*_{ij}u_{ij}\theta_j +  c.c.\nonumber\\
&=&\sum_i\sum_{j\sim i}\theta_i v^*_{ij} u^*_{ij}+ c.c.\ ,
\label{varmorseuv}
\end{eqnarray}
where $u_{ji}=u^*_{ij}$ and $v_{ji}=-v_{ij}$ was used to rewrite the
second term.
On the other hand\equ{varmorse} together with\equ{f} imply
\begin{equation}
\Delta V=-\sum_i (f^*_i[U]\theta_i + c.c.)
\label{varm}
\end{equation}
Comparing\equ{varm} with\equ{varmorseuv} one obtains 
\begin{equation}
f_i[u,v]=-\sum_{j\sim i} v_{ij} u_{ij}\ .
\label{fuv}
\end{equation}  
We can compute the linear operator $M_i[U,c]$ in analogous fashion by
considering the variation of $f_i[u,v]$ under infinitesimal 
transformations of the form\equ{varuv}. One gets
\begin{eqnarray}  
\left. s f_i[u,v]\right|_{\omega=0}=\sum_{j\sim i} C_i
(|v_{ij}|^2-|u_{ij}|^2) +u^2_{ij} C_j -v^2_{ij} C^*_j \ ,
\label{sf}
\end{eqnarray}
where the Grassmann variables $C_i,C^*_i$ are defined by the
decomposition 
\begin{equation}
c_i=C_i\tau_+ - C_i^*\tau_-
\label{C}
\end{equation}
Similarly decomposing $\bar c$ as
\begin{equation}
\bar c_i=\bar C_i\tau_+ + \bar C_i^*\tau_-\ ,
\label{Cbar}
\end{equation}
one obtains for the quadratic form
\begin{eqnarray}
\sum_i\Tr\bar c_i M_i[U,c]&=&\sum_i\left.\Tr\bar c_i s
F_i[U]\right|_{\omega=0}\nonumber\\
&=&-\sum_i(\left.\bar C_i s f_i[u,v]\right|_{\omega=0}
-c.c.)\nonumber\\
&=&\sum_i\sum_{j\sim i}\bar C_i(|u_{ij}|^2-|v_{ij}|^2) C_i +\bar
C_i v^2_{ij} C^*_j - \bar C_i u^2_{ij} C_j- c.c.\ .
\label{quadratic}
\end{eqnarray}
Here ``complex conjugation'' for Grassmann variables is the
substitution  $C,\bar C\leftrightarrow C^*,\bar C^*$ at each site.  

Using\equ{C} and\equ{Cbar} the interaction with the real scalar field
$\rho_i$ in\equ{linSGF} is written
\begin{equation}
\sum_i \rho_i\Tr\tau_0[\bar c_i,c_i]=-\sum_i \rho_i (\bar C_i C^*_i
+\bar C^*_i C_i)\ .
\label{rhoint}
\end{equation}
Defining the two complex $N\times N$ matrices with entries,
\begin{eqnarray}
A_{ij}[u]&=&u^2_{ij}+\delta_{ij}\sum_{k\sim i} (1-2|u_{ik}|^2) 
\nonumber\\
B_{ij}[v,\rho]&=& \delta_{ij}\rho_i- v^2_{ij}\ ,
\label{defmat}
\end{eqnarray}
the integration of the Grassmannian variables in the gauge fixing
part of the action\equ{linSGF} results in a weight proportional to
\begin{equation}
\det {\cal M}[u,v,\rho]\ , 
\label{FP}
\end{equation}
for the remaining bosonic functional integral. The $2N\times 2N$ complex
matrix ${\cal M}$ is 
\begin{equation}
{\cal M}[u,v,\rho]=\left(\begin{array}{c|c} A[u] & B[v,\rho]\\
                         \hline
B^\dagger[v,-\rho] & A^\dagger[u]\end{array}\right)\ .
\label{M}
\end{equation}                  
Note the dependence on the auxiliary field $\rho$ in\equ{M}. For
purely imaginary $\rho_i$ the matrix ${\cal M}$ is hermitian and its
eigenvalues (and thus its determinant) are real. The weight\equ{FP} of
a given bosonic configuration with real $\rho_i$  is however  generally
complex. One can also corroborate that $A[u_{ij}=1]$ is the lattice
Laplacian with exactly {\it one} vanishing eigenvalue on a periodic
lattice.  
${\cal M}[1,0,0]$ thus has exactly two vanishing eigenvalues
corresponding to the two zero-modes 
of this vacuum configuration that were found in section~IV. The same 
reasoning shows that, $\det{\cal M}[u,v,0]$ in fact vanishes for any
pure gauge configuration. 

Collecting these results, the gauge-fixing weight ${\cal W}[u,v,\rho;\alpha]$
of a given link configuration can be written
\begin{equation}
{\cal W}[u,v,\rho;\alpha]=\det{\cal M}[u,v,\rho] \prod_{sites}
e^{-\rho_i^2/(4\alpha)} \
P\left(\frac{f_i[u,v]}{\sqrt{\alpha}}\right)\ ,
\label{gfweight}
\end{equation}
where the local weight $P(x)$ depends on whether the  $SU(2)$-LGT is
partially gauge fixed to the abelian $U(1)$- or the discrete $Z_2$- structure group
\begin{equation}
P(x) = \left\{\begin{array}{cc} e^{-|x|^2} & \text{ for }
U(1)\nonumber\\
{\cal P}({\rm Im}\, x)\text{ of\equ{P} } & \text{ for }Z_2
\end{array}\right.
\label{U1P}
\end{equation}
The expectation value of operators ${\cal O}[u,v,\rho]$ that only
depend on the link variables and the auxiliary field $\rho$ can now be
found by (numerically) 
evaluating the remaining  bosonic integrals in
\begin{equation}
\vev{{\cal O}[u,v,\rho]}=\int \prod_{links} d^2 u_{ij} d^2
v_{ij} \delta\left(1-|u_{ij}|^2-|v_{ij}|^2\right) \prod_{sites}
d\rho_i {\cal O}[u,v,\rho] {\cal W}[u,v,\rho;\alpha] e^{-S_{\rm inv.}[u,v]}\ .
\label{bosonint}
\end{equation}

\end{document}